\documentclass[journal]{IEEEtran}

\usepackage[T1]{fontenc} 
\usepackage{cite}
\usepackage{algpseudocode}
\usepackage{subfigure}
\usepackage{graphicx}
\usepackage{amsmath}
\usepackage{mathtools}
\usepackage{textcomp}
\usepackage{etoolbox}
\usepackage{comment}
\usepackage{enumitem}
\usepackage{booktabs}
\usepackage{multirow} 
\usepackage{epstopdf}
\usepackage{bm}
\newcommand\bolden[1]{{\boldmath\bfseries#1}}
\usepackage{gensymb}
\usepackage{array}
\usepackage{wrapfig}
\usepackage{algorithm}
\usepackage[flushleft]{threeparttable}
\usepackage{xcolor}
\usepackage{soul}

\newcommand{\innermid}{\nonscript\;\delimsize\vert\nonscript\;}
\newcommand{\activatebar}{
  \begingroup\lccode`\~=`\|
  \lowercase{\endgroup\let~}\innermid 
  \mathcode`|=\string"8000
}
\usepackage{amsthm,amssymb}
\usepackage{setspace}
\usepackage{rotating,booktabs}
\newcommand{\subparagraph}{}
\newcommand{\etal}{\textit{et al.}}
\usepackage[compact]{titlesec}
\usepackage{hyperref}
\hypersetup{
    colorlinks=false,
}
\usepackage{bm}
\usepackage{tikz}
\usetikzlibrary{shapes,arrows}
\def\tablename{TABLE}
\begin{document}
\title{Learning to Compress Videos without Computing Motion}

\author{Meixu Chen,~\IEEEmembership{Student Member,~IEEE,}
        Todd Goodall, 
         Anjul Patney, and Alan C. Bovik,~\IEEEmembership{Fellow,~IEEE}% <-this % stops a space
\thanks{M. Chen and A. C. Bovik are with the Department of Electrical and Computer Engineering, University of Texas at Austin, Austin, USA (e-mail: chenmx@utexas.edu; bovik@ece.utexas.edu). T. Goodall was with Facebook Reality Labs when this work was conducted, and now he is with Apple, Inc (email: beyondmetis@gmail.com). A. Patney was with Facebook Reality Labs when this work was conducted, and now he is with Nvidia. (email: rwebb@fb.com). }}

% \markboth{Journal of \LaTeX\ Class Files,~Vol.~14, No.~8, August~2015}%
% {Shell \MakeLowercase{\textit{et al.}}: Bare Demo of IEEEtran.cls for IEEE Journals}

\maketitle

\begin{abstract}
Video has become an increasingly important part of our daily digital communication. With the development of higher resolution contents and displays, its significant volume poses significant challenges to
the goals of acquiring, transmitting, compressing and displaying
high quality video content. 
In this paper, we propose a new deep learning video compression architecture that does not require motion estimation, which is the most expensive element of modern hybrid video compression codecs like H.264 and HEVC. Our framework exploits the regularities inherent to video motion, which we capture by using displaced frame differences as video representations to train the neural network. In addition, 
we propose a new space-time reconstruction network based on both an LSTM model and a UNet model, which we call LSTM-UNet. The combined network is able to efficiently capture both temporal and spatial video information, making it highly amenable for our purposes. 
The new video compression framework has three components: a Displacement Calculation Unit (DCU), a Displacement Compression Network (DCN), and a Frame Reconstruction Network (FRN), all of which are jointly optimized against a single perceptual loss function. The DCU \textit{removes the need for  motion estimation found in hybrid codecs,} and is  less expensive. In the DCN, an RNN-based network is utilized to compress displaced frame differences as well as retain temporal information between frames. The LSTM-UNet is used in the FRN to learn space time differential representations of videos. 
Our experimental results show that our compression model, which we call the MOtionless VIdeo Codec (MOVI-Codec), learns how to efficiently compress videos without computing motion. Our experiments show that MOVI-Codec outperforms the Low-Delay P (LDP) \textit{veryfast} setting of the video coding standard H.264 and exceeds the performance of the modern global standard HEVC codec, using the same setting, as measured by MS-SSIM, especially on higher resolution videos. In addition, our  network  outperforms  the  latest  H.266 (VVC) codec at higher bitrates, when assessed using MS-SSIM, on high resolution videos. The MOVI-Codec project page can be found at \url{https://github.com/Meixu-Chen/MOVI-Codec}. 
\end{abstract}

\IEEEpeerreviewmaketitle
\section{Introduction}
\IEEEPARstart{V}{ideo} traffic is predicted to reach 82 percent of all
consumer Internet traffic by 2021 \cite{index2017cisco}, and to continue this rapid growth even further. The increasing share of video in Internet traffic is being driven by several factors, including the great diversity and extraordinary popularity of streaming and social media services, the rise of video teleconferencing and online video education (accelerated by the Coronavirus Crisis), and significant increases in video resolution. Indeed, it is estimated that by 2023, two-thirds of installed flat-panel television sets will be UHD, up from 33 percent in 2018 \cite{cisco}. Given significant strains on available bandwidth, it is crucial to 
continue and greatly accelerate the evolution of video compression systems. 
%  of not only traditional ones like TV and computer, but also newer applications like Virtual Reality and Augmented Reality \cite{chen2019study}

Traditional video compression codecs, like H.264,  HEVC and the latest VVC/H.266 process videos through a sequence of hand-designed algorithms and modules, including block motion estimation, and local decorrelating decompositions like the Discrete Cosine Transform (DCT). Although the component modules of modern hybrid codecs have been carefully designed over several generations, the overall codecs have not been globally optimized other than by visual examination or post-facto objective measurement of results, typically by the highly fallible PSNR \cite{wang2009mean}. Naturally, one could expect the performances of video codecs to be improved by collective, end-to-end optimization. Because of their tremendous ability to learn efficient visual representations, deep learning models are viewed as highly promising vehicles of developing alternative, globally optimal video codecs, and a variety of deep learning based image compression architectures have been proposed \cite{toderici2015variable, toderici2017full, balle2018variational, agustsson2017soft, balle2019end, johnston2018improved, theis2017lossy, rippel2017real, agustsson2019generative, mentzer2018conditional, minnen2018joint, patel2019deep, lee2018context, blau2019rethinking, mentzer2019practical, chen2017deepcoder}. These new models have deployed Convolutional Neural Networks (CNN), Recurrent Neural Networks (RNN), autoencoders, and Generative Adverserial Networks (GAN) yielding rate-distortion efficiencies that are reportedly comparable to those of traditional image compression codecs like JPEG, JPEG 2000, and BPG. Encouraged by these advances, several authors have devised deep video compression models that suggest the considerable promise of this general approach. Wu \textit{et al.} \cite{wu2018video} proposed the first end-to-end trained deep video codec, using a hierarchical frame interpolation scheme. A block-based deep video compression codec was proposed by Chen \etal \cite{chen2017deepcoder}. 
Lu \textit{et al.} proposed an end-to-end video compression model (DVC) \cite{lu2019dvc} which replaces each component of the traditional hybrid video codec with a deep learning model, which are jointly trained as a global hybrid architecture against a single loss function. Another hierarchical video compression architecture called HLVC (Hierarchical Learned Video Compression) was proposed by Yang \textit{et al.}\cite{yang2020learning}. 

In both traditional video codecs and recent deep learning-based ones, motion estimation and compensation has occupied a significant portion of the system resources.  Motion estimation requires an expensive search process that we avoid, by instead training the network to efficiently represent the residuals between each current frame and a set of spatially-displaced neighboring frames. Computing a set of frame differences, even over many displacement directions is much cheaper than effective search processes. Moreover, while the statistics of motion are generally not regular, the intrinsic statistics of frame differences exhibit strong regularities\cite{soundararajan2012video}, including those of differences between spatially displaced frames \cite{lee2021space}. The strong internal structure of these frame differences makes them easier to efficiently represent in a deep architecture.

Our idea is inspired by the way the human vision system processes natural time-varying images. Many studies have produced strong evidence suggesting that the early stages of the vision system are primarily implicated in reducing redundancies in the sensed input visual signals \cite{atick1990towards,attneave1954some}. Indeed, much of early visual processing along the retino-cortical pathway appears to be devoted to process of spatial and temporal decorrelation\cite{dong1995temporal, rucci2015unsteady, chichilnisky2002functional, engbert2006microsaccades, poletti2016compact, olshausen1996emergence}. We have found that sets of spatially displaced frame differences, which are space-time processes, supply a rich and general way to exploit space-time redundancies\cite{soundararajan2012video,lee2021space}. Importantly, our idea is also related to recent theories of the role of microsaccades in human visual information processing\cite{rucci2015unsteady, chichilnisky2002functional, engbert2006microsaccades, poletti2016compact}. Microsaccades create small spatial displacements of visual field from moment to moment. While microsaccades have been theorized to play roles in avoiding retinal saturation, maintaining accurate fixation in the presence of drifts, and preserving the perception of fine spatial details \cite{poletti2016compact}, they are more recently thought to play an important role in efficiently representing locally changing and shifting space-time visual information \cite{rucci2015unsteady, chichilnisky2002functional, engbert2006microsaccades, poletti2016compact}. We believe that micro-saccadic eye movements deployed by the human eye have adapted to the local regularities induced by small spatial displacements over time, in order to achieve more efficient visual (neural) representations. This has inspired us to, in like manner, trained a deep coder-decoder network to compress videos using regular displaced residual representations as inputs.

By capturing displaced frame differences from a large database of videos, and feeding them into a deep space-time coding-decoding network, we have formulated a new breed of deep video compression algorithms that are motion computation free, statistically motivated, and have perceptual relevance.  The contributions of this work can be summarized as follows: 
\begin{itemize}
    \item We innovate the use of displaced frame differences to capture efficient representations of structures induced by motion. 
    \item Our method avoids the computational overhead of motion estimation and motion compensation.
    \item A combined LTSM-UNet efficiently captures both spatial and temporal information which it uses to recreate video frames from the abstracted video code. 
    \item The entire video compression system is collectively jointly optimized using a single loss function.
\end{itemize}
Our results show that video compression can be efficiently accomplished without explicitly computing motion predictions. We trained the new MOVI-Codec architecture end-to-end on the Kinetics-600 dataset and the Vimeo-90K dataset, using a single perceptual loss function (MS-SSIM), and tested it on the UVG dataset, the VTL dataset, and the HEVC Standard Test Sequences (Class B, Class C, Class D, and Class E). Our experimented results show that our new model outperforms the widely used video codec H.264 in LDP \textit{veryfast} setting, and exceeds the performance of  the latest standard video codec H.265 using the same setting. In addition, our  network  outperforms  the  latest  H.266 (VVC) codec at higher bitrates, as assessed by the perceptually relevant MS-SSIM algorithm, on high resolution videos.

The rest of the paper is organized as follows. 
Section \ref{section2} briefly introduces current progress on learning-based methods for image/video compression and motion estimation.
Section \ref{section3} describes details of the architecture and training protocol of the new MOVI-Codec model.
Section \ref{section4}  discusses the experiments we conducted and their outcomes, along with a data analysis along several dimensions. Section \ref{section6} concludes the paper with a discussion of future research directions.

\section{Related works}
\label{section2}
\subsection{Deep Image Compression}
 A variety of standardized image compression engines have been proposed over the years to meet the needs of increasingly picture-centric technologies. JPEG algorithm \cite{wallace1992jpeg}, and later challengers JPEG 2000 \cite{skodras2001jpeg}, BPG \cite{bellard2015bpg}, and VP9 \cite{mukherjee2013latest}. These methods have proven to be quite practical, and in the case of JPEG, ubiquitous. Yet they are all handcrafted, highly modularized without the benefit of collective optimization of all their elements. Each of these standards maps pixels to a less correlated representation, regardless of the attributes of the input image. These transformed values are then non-uniformly quantized, typically with reference to a human visual sensitivity model. 
 
 A variety of authors have recognized the potential of deep learning to advance progress on the image compression problem (a still timely goal given the senectitude of the prevailing JPEG standard), and many learning-based architectures have been devised \cite{toderici2015variable, toderici2017full, balle2018variational, agustsson2017soft, balle2019end, johnston2018improved, theis2017lossy, rippel2017real, agustsson2019generative, mentzer2018conditional, minnen2018joint, patel2019deep, lee2018context, blau2019rethinking, mentzer2019practical}. Given that Convolutional Neural Networks (CNN) \cite{lecun1995convolutional} were the first deep learning models to obtain standout performance on image analysis problems, it was natural that it be the first deep architecture to be applied to learning-based image compression. Ball{\'e} \etal \cite{balle2019end} proposed a CNN-based image compression framework that was optimized end-to-end, which was shown to outperform JPEG2000 with respect to both MS-SSIM and PSNR image quality measures. Their framework was later extended by incorporating a hyperprior to capture
spatial dependencies in the latent representation for entropy estimation \cite{balle2018variational}. In \cite{minnen2018joint}, Minnen \textit{et al.} further enhanced the entropy model, by combining autoregressive and hierarchical priors to exploit the probabilistic
structure in the latents. The resulting model was reported to outperform BPG with respect to both PSNR and MS-SSIM. Another architecture favored for learning-based image compression are Recurrent Neural Networks (RNN), because of their ability to exploit representative memories. Long Short-Term Memory (LSTM) models were proposed \cite{hochreiter1997long} to address the vanishing gradient problem of RNNs. Toderici \etal \cite{toderici2015variable, toderici2017full} was the first to deploy a deep RNN-based architecture for image compression by utilizing a scale-additive framework. This architecture allows for variable bit rates and only needs to be trained once. The authors also presented results using different types of RNNs, including LSTM, associative LSTM and a hybrid of a Gated Recurrent Unit (GRU) \cite{cho2014properties} and a ResNet, reporting that the performance of the model was better than JPEG. Generative Adversarial Networks (GAN) have been applied in several learning-based image compression models. Early on, Rippel \etal \cite{rippel2017real} proposed a GAN-based image compression framework that they claim outperformed all existing codecs with respect to MS-SSIM, while being lightweight and deployable. In \cite{agustsson2019generative}, a GAN framework is presented to build an extreme image compression system which the authors report as achieving state-of-the-art performance, especially at very low bit rates, based on a user study.
\subsection{Deep Video Compression}
It is natural to also consider learning-based methods for video compression \cite{chen2017deepcoder, wu2018video, yang2020learning, lu2019dvc,  habibian2019video, rippel2019learned, chen2019learning, cheng2019learning}. 
Wu \textit{et al.} \cite{wu2018video} proposed a video compression architecture based on the idea that video compression is repeated image compression. They define two types of frames: key frames and other frames. Key frames are compressed using an RNN-based image compression network \cite{toderici2017full}, while the other frames are interpolated in a hierarchical manner. Another hierarchical video compression architecture, called Hierarchical Learned Video Compression (HLVC), was proposed by Yang \textit{et al.}\cite{yang2020learning}. In this method, there are three quality layers: an image compression layer, a Bi-Directional Deep Compression (BDDC) layer, and a Single Motion Deep Compression (SMDC) layer. In an attempt to match the pipeline structure of hybrid codecs, Lu \textit{et al.} proposed an end-to-end video compression model (DVC) \cite{lu2019dvc} that replaces each traditional hybrid component, with deep learning models, then jointly optimized all the components against a single loss function. This work was further extended to two models, a lightweight version called DVC\_Lite, and an advanced version called DVC\_Pro, by adjusting various components of the architecture. Later, Habibian \etal \cite{habibian2019video} proposed a deep generative model for video compression using an autoregressive prior to conduct entropy coding. Generally, all learning-based video compression models implement traditional block-based motion estimation or optical flow, both of which have a high computational overhead. 

The most related work to ours is \cite{cheng2019learning}, whereby an interpolation loop is used as an alternative to motion estimation/compensation. However, the frame interpolation network still requires training, which adds to the complexity of the overall method. 

\subsection{Motion Estimation and Motion Compensation}
Motion estimation (ME) and motion compensation (MC) are crucial components in modern hybrid video codecs. These are used to exploit the temporal redundancy of video frames via inter-frame prediction. In traditional hybrid video codecs like H.264 and H.265, video frames are first partitioned into blocks, then motion vectors (MV) associated with each block are estimated with respect to predictions of neighboring reference frames via expensive block search methods, which is the most intensive aspect of video compression. A few deep learning methods have been proposed to solve the ME problem. For example, Choi \textit{et al. }\cite{choi2017new} trained a CNN to measure the similarity of  pairs of image patches and used this to estimate MVs. However, this method still requires a search process to find the best match. In \cite{choi2019deep}, the authors  developed a CNN that was trained to conduct both uni- and bi-directional ME, using separate networks so that motion information need not be transferred from the encoder to the decoder. The CNN does require two frames from the decoded picture buffer and their temporal indices as inputs, which it uses to produce filter coefficients that synthesize patches of a new frame, which is then used to predict the current frame. A drawback of this approach is that it requires the CNN to be resident at both the encoder and the decoder, which reduces decoding efficiency. 

Another popular alternative to block matching algorithms are optical flow routines, which seek to obtain a dense vector field  mapping the movements of pixel. A variety of deep learning based optical flow estimation methods have been proposed to reduce the  computational overhead of dense optical flow vectors \cite{ranjan2017optical}. FlowNet \cite{dosovitskiy2015flownet} showed that it was possible to train a network from two input images to predict optical flow while matching or exceeding the accuracies of traditional methods. Later improvements introduced a stacked architecture that included warping of the second image via intermediate optical flow estimates, and a sub-network specialized to predict small motions \cite{ilg2017flownet}. Other approaches have tried to combine networks with traditional methods. Ranjan \textit{et al.} \cite{ranjan2017optical} proposed such a network called SpyNet, which adopted a traditional coarse-to-fine computational hierarchy using a spatial pyramid. Later, another network competitive with FlowNet2 was proposed, called LiteFlowNet \cite{hui2018liteflownet}, but with a significantly decreased model size. Our approach avoids even these methods of deep flow computation, by instead feeding the network a set of directional inter-frame residuals containing adequate information for the network to seek the most efficient perceptual representation.

\section{Proposed Method}
\label{section3}
\subsection{Framework}

\begin{figure} [!t]
\centerline{
\includegraphics[width=1\columnwidth]{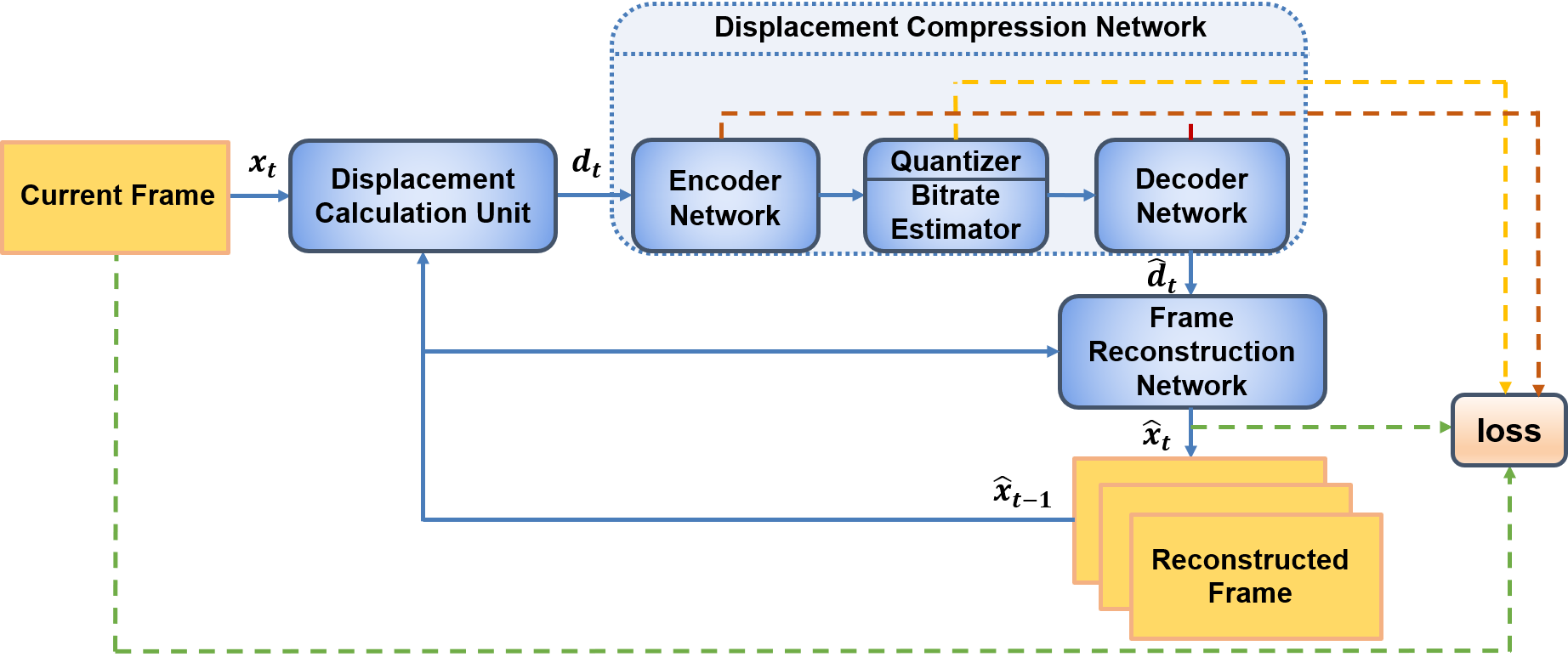}
}
\caption{The overall network architecture of MOVI-Codec, which consists of three components: a Displacement Calculation Unit, a Displacement Compression Network and a Frame Reconstruction Network.} 
\label{fig:network}
\end{figure}

Figure \ref{fig:network}  exemplifies the flow of our deep video compression network. A current frame is input to the network, along with multiple displaced frame differences from adjoining,  previously coded and then decoded frames (lower part of figure). This is similar to the classic hybrid coding loop, which also includes the decoder as part of the encoder loop, to reduce reconstruction errors. The key components in our network is: Displacement Calculation Unit (DCU), Displacement Compression Network (DCN), and Frame Reconstruction Network (FRN). The details of each key component in our network will be discussed in the following sections. 

The flow of our network is: Given an input video with frames $x_1, x_2, ..., x_T$, for every frame $x_t$, displaced frame differences between the current frame $x_t$ and previous reconstructed frame $\hat{x}_{t-1}$ are calculated via the DCU, after which the displaced frame differences $d_t$ are input into the DCN. The DCN compresses the incoming displaced frame differences which are used to capture statsitical redundancies. An illustration of displaced frame differences, i.e. differences between spatially displaced frames, is shown in Figure \ref{fig:displacement}. Given a compressed output $\hat{d_t}$ from the DCN, FRN uses the reconstructed displaced frame differences $\hat{d_t}$ and the reconstructed previous frame $\hat{x}_{t-1}$ to reconstruct a current frame $\hat{x_t}$. Every frame is processed following this except for the first frame. The first frame $x_1$ is processed differently as it does not have previous reconstructed frame. As a result, an all-zero image is chosen as its previous reconstructed frame and it is otherwise processed the same as other frames. Pseudo code of the flow is shown in Algorithm \ref{alg:procedure}. By using this architecture, we are able to reconstruct the videos without the use of motion. 

\begin{algorithm}
\caption{Flow of MOVI-Codec for an Input Video}
\label{alg:procedure}
\bolden{$x_1$ to $x_T$:} video frames. \\
\bolden{$\hat{x}_0$:} previous reconstructed frame for $x_1$.\\
\bolden{$d_t$, $\hat{d}_t$:} displaced frame differences and corresponding reconstructed ones, respectively.\\
\bolden{$d_1$, $\hat{d}_1$:} displaced frame differences between $x_1$ and $\hat{x}_0$, and corresponding reconstructed ones, respectively.
\begin{algorithmic}[1]
    \Procedure{MOVI-Codec}{}
    \For{$t$ in 1 to $T$}
        \If{$t$ is 1}
            \State $\hat{x}_0$ = all zero frame
            \State $d_1$ $\gets$ DCU($x_1$, $\hat{x}_0$)
            \State $\hat{d}_1$ $\gets$ DCN($d_1$)
            \State $\hat{x}_1$ $\gets$ FRN($\hat{d}_1, \hat{x}_0$)
        \Else
            \State $d_t \gets$ DCU($x_t$, $\hat{x}_{t-1}$)
            \State $\hat{d}_t \gets$ DCN($d_t$)
            \State $\hat{x}_t \gets$ FRN($\hat{d}_t, \hat{x}_{t-1}$)
        \EndIf
    \EndFor
    % \State \textbf{return} $\hat{x}_{t}$
    \EndProcedure
\end{algorithmic}
\end{algorithm}

\subsection{Displacement Calculation Unit (DCU)}
\begin{figure} [!t]
\centerline{
\includegraphics[width=0.8\columnwidth]{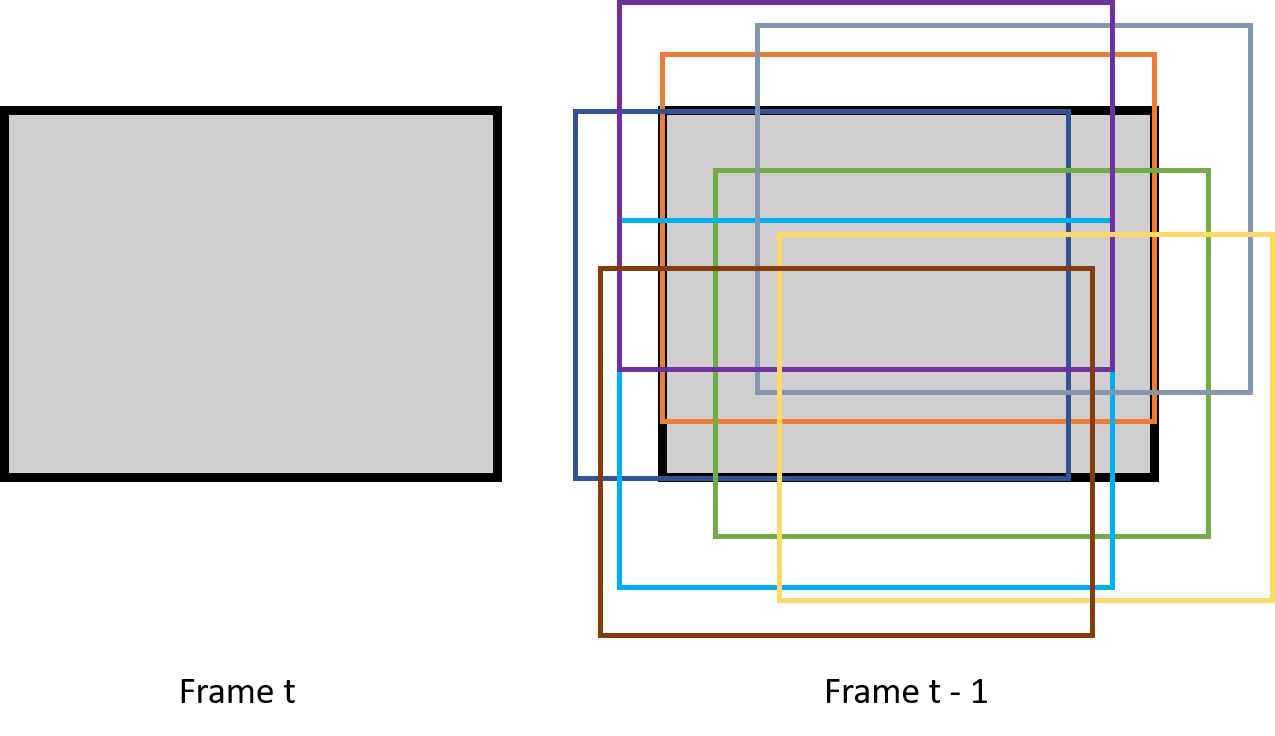}
}
\caption{Concept of displaced frame differences, showing a frame $t$ and previous frame $t - 1 $, and multiple spatially displaced versions of frame $t - 1$.}
\label{fig:displacement}
\end{figure}

The DCU removes the need for any kind of motion vector search. Instead, it allows the DCU network to learn to optimally represent time-varying images as sets of spatially displaced frame differences. Given a video with $T$ frames ${x_1, x_2, ..., x_T}$ of width $w$ and height $h$, two directional (spatially displaced) temporal differences are computed between each pair of adjacent frames, as shown in Figure \ref{fig:displacement}. In the DCU, the inputs are a current frame $x_t$ and the reconstructed previous frame $\hat{x}_{t-1}$. Then, at each spatial coordinate $(i, j)$, a set of spatially displaced differences is calculated as: 
    \begin{equation}
        d_H(i,j)_{t} = x_t(i,j) - \hat{x}_{t-1}(i, j - s),
    \end{equation}
    \begin{equation}
        d_V(i,j)_{t} = x_t(i,j) - \hat{x}_{t-1}(i - s, j),
    \end{equation}
where $s = 0, \pm{3}, \pm{5}, \pm{7}$ in our experiment.
The set of 13 displaced frame differences (residuals) is then fed into the Displacement Compression Network, which delivers as output the reconstructed set of displaced residuals $\hat{d_t}$. As mentioned in Section \ref{section2}, the statistics of non-displaced frame differences have been observed to be nicely regular. As shown in \cite{lee2021space}, the statistics of displaced frame differences are also highly regular, and more so in the direction of local motion. This makes them good video representations to learn to exploit space-time redundancies, while avoiding the computational burden of motion estimation and compensation. Although the range of motion between frames can be larger than our largest choice of displacement, larger motions can be captured by various combinations of our set of displacements.
\subsection{Displacement Compression Network (DCN)}
\subsubsection{Framework}
After a set of 13 displaced frames are generated from the Displacement Calculation Unit, they are fed into the Displacement Compression Network, where each displacement occupies three channels (RGB), hence the overall input to the DCN comprises 39 channels. The compression network comprises four parts, displacement encoder, displacement decoder, hyper encoder, and hyper decoder. Displacement encoder takes the displaced frame differences calculated from DCU and generates the latent representation $y_t$ using several convolutional layers and convolutional LSTM layers similar to other deep learning-based compression architectures \cite{wu2018video, toderici2017full}. LSTM \cite{xingjian2015convolutional} as a special RNN structure has proven stable and
powerful for modeling long-range dependencies in sequence modeling. The
major innovation of LSTM is its memory cell which keeps accumulating the
state information. As a result, it helps hold the spatio-temporal information provided by displaced frame differences generated by DCU. The hyper autoencoder uses $y_t$ as input to generate side information, which is then used to better compress quantized latent representation $\hat{y}_t$. Finally, the reconstructed $\hat{d}_t$ is generated using $\hat{y}_t$. The detailed processing flow of the hyper autoencoder is explained in later sections.

\begin{figure*} [!t]
\centerline{
\includegraphics[width=1.6\columnwidth]{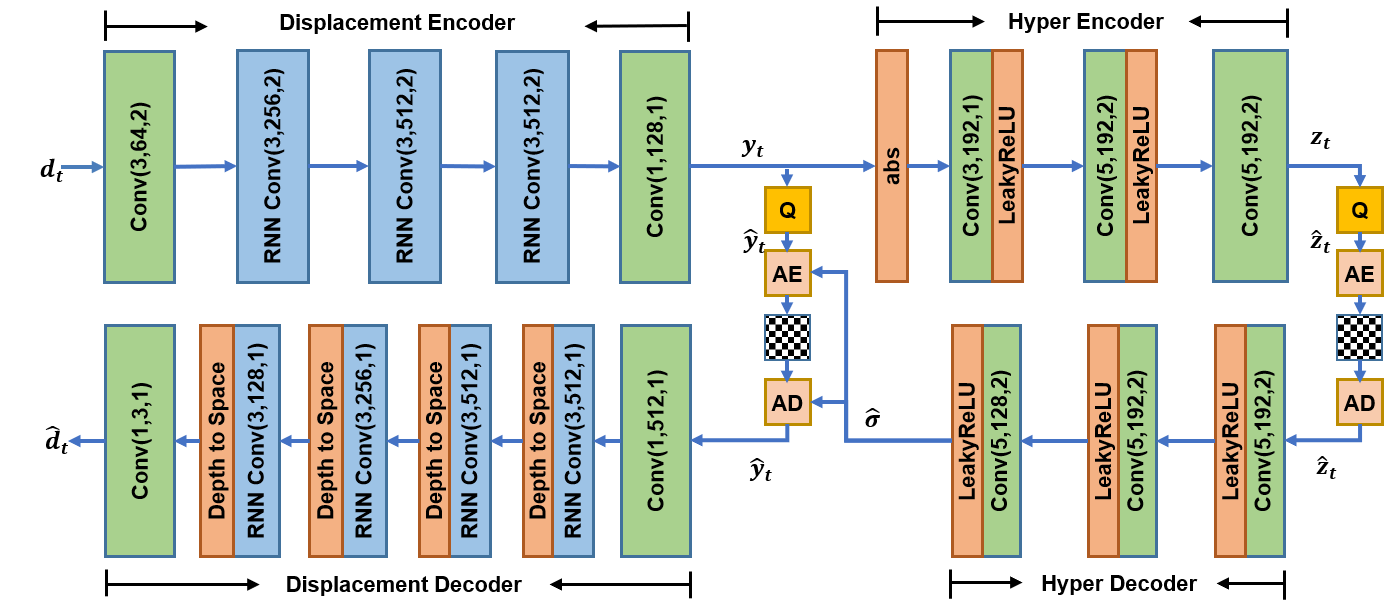}
}
\caption{Flow diagram of the Displacement Compression Network. The left side shows the displacement autoencoder architecture, and the right side corresponds to the hyperprior autoencoder architecture. Q represents quantization, and AE, AD represent arithmetic encoder and arithmetic decoder, respectively. Conv(3,64,2) represents the convolution operation with  kernel size of 3x3, 64 output channels and a stride of 2.}
\label{fig:compression}
\end{figure*}
\subsubsection{Quantizer}
Traditional quantization  inevitably produces zero gradients during backpropagation (BP) which halts network training. Our network deploys BP via stochastic gradient descent, which requires differentiability of all network elements. Hence, we implemented a modified quantizer as in \cite{balle2018variational}, as follows, where $\hat{y}$ is the binarization of the latent representation of displaced frame differences, which lie between -1 and 1, and $\epsilon$ represents quantization noise: 

    \begin{equation}
        \hat{y} = y + \epsilon \in{-1,1}
    \end{equation}
    \begin{equation}
        \epsilon \sim
        \begin{cases}
        1 - y & \text{with probability }  \frac{1 + y}{2} \\
        -y - 1 & \text{with probability }  \frac{1 - y}{2}
        \end{cases}.
    \end{equation}
Following quantization, the size of $\hat{y}$ is $\frac{H}{16} \times \frac{W}{16} \times C$, where $H$ and $W$ are the height and width of the frame,  and $C$ is the number of channels of the last convolution layer in the displacement encoder. In our architecture, $C =128$, as shown in Figure 3.
\subsubsection{Entropy Coding}
To estimate the entropy of the compressed codes $H(\hat{y})$, where $\hat{y}$ is the quantized latent representation of $y$, we adopted the hyper-prior scheme proposed by Ball{\'e} \textit{et al.} \cite{balle2018variational}, where they use an additional set of random variables $\hat{z}$ to capture the spatial dependencies and model the latent representations $\hat{y}$ as Gaussian distribution as follows:
\begin{equation}
    p_{\hat{y}|\hat{z}} (\hat{y}|\hat{z}) \sim \mathcal{N}(\mu, \sigma), 
\end{equation}
where $p_{\hat{z}} (\hat{z})$ is modeled using the factorized entropy model \cite{balle2019end}.

The hyperprior autoencoder architecture is indicated by Hyper Encoder and Hyper Decoder in Figure \ref{fig:compression}, which is responsible for estimating the parameters of the Gaussian model used for entropy coding. After the displacement encoder encoded the input set of displaced frame differences $d_t$, the resulting latent representation $y_t$ with spatially varying standard deviations is fed into the hyper encoder, which summarizes the distribution of standard deviations in the latent representation $z_t$. After quantization and arithmetic coding, the quantized $\hat{z}_t$ is transmitted as side information. The hyper decoder uses the quantized $\hat{z}_t$ as input to obtain Gaussian model parameter $\hat{\sigma}$ ($\mu$=0 in our implementation). During modeling training, the Gaussian model parameters can be used to calculate $p_{\hat{y}_t}$ and then estimate $H(\hat{y}_t)$ to guide model optimization. While during model validation and/or testing, the Gaussian model can be used to calculate the cumulative distribution function (CDF) of  $\hat{y}_t$ and then guide the arithmetic encoding and decoding of $\hat{y}_t$, which could further losslessly compress $\hat{y}_t$ to bitstream. 
% The detailed procedure of entropy coding is shown in Algorithm 2. %%% put a algo chart here

\subsection{Frame Reconstruction Network (FRN)}
\begin{figure} [!t]
\centerline{
   \label{LSTM-UNet}
\includegraphics[width=0.7\columnwidth]{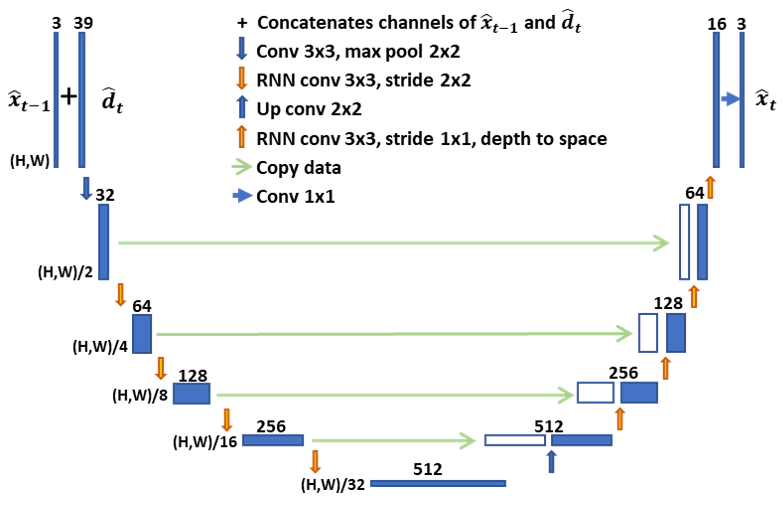}}
\caption{LSTM-UNet architecture used in Frame Reconstruction Network. }
\label{fig:unets}
\end{figure}

 Figure \ref{fig:unets} shows the structure of the Frame Reconstruction Network (FRN). The FRN uses the reconstructed displaced frame differences $\hat{d_t}$ and the reconstructed previous frame ${\hat{x}_{t-1}}$ as the model input to reconstruct the current frame. The architecture of FRN incorporates Convolutional LSTM (C-LSTM) blocks into a UNet  architecture. The UNet  architecture, which is an encoder-decoder style network with skip connections, makes it possible to extract and represent meaningful descriptors over multiple image scales. However, without modification, the UNet architecture cannot account for temporal relationships between frames of video data, which are deeply relevant to the efficiency of video compression. The C-LSTM is a convolutional version of the original LSTM, which replaces the matrix multiplication operation of the traditional LSTM with convolutions. It is quite useful for analyzing temporal image sequences, where the C-LSTM layers act as a temporal buffer and capture the long-short dependency of previously processed displaced frame differences. By introducing C-LSTM blocks into the UNet architecture, the FRN is able to process evolving frame properties over multiple scales, by relating compact representations of them in the C-LSTM memory units, leading to better reconstructed frame quality and higher compression rates.

\subsection{Training Strategy}

We modeled the loss function considering the rate-distortion trade-off as follows:
\begin{equation}
\label{loss}
    \begin{split}
L  &  = D + \lambda R \\
& = [{D_1(x_t, \hat{x}_t)} + \beta D_2(d_t, \hat{d}_t)] + \lambda [H(\hat{y}_t) + H(\hat{z}_t)],
    \end{split}
\end{equation}
where $D$ and $R$ represent the distortion and rate, respectively. $\lambda$ controls the trade-off between the number of bits and distortion. $D_1$ denotes the distortion between the input frame $x_t$ and reconstructed frame $\hat{x}_t$ measured by MS-SSIM or MSE, and $D_2$ denotes the distortion between displaced frame differences $d_t$ and the reconstructed displaced frame differences $\hat{d}_t$ measured by MSE. $\beta$ controls the trade-off between the perceptual distortion $D_1$ and the pixel-to-pixel distortion $D_2$. $H(\cdot)$ represents the bitrates for encoding the latent representations $\hat{y}$ and $\hat{z}$ estimated by the hyperprior autoencoder. 

To leverage multi-frame information using our RNN-based codec structure, we update the network parameters every set of $N$ frames during model training, using the loss function in Equation \ref{loss} but modified as a sum of losses over the $k$th set of the $N$ frames indexed $x_{{t_k}+1}, ..., x_{{t_k}+N}$:
\begin{multline}
    % \begin{split}
L_k  = \frac{1}{N}\sum_{n=1}^{N}[{D_1(x_{{t_k}+n}, \hat{x}_{{t_k}+n})} + \beta D_2(d_{{t_k}+n}, \hat{d}_{{t_k}+n})] \\ +  \lambda [H(\hat{y}_{{t_k}+n}) + H(\hat{z}_{{t_k}+n})].
    % \end{split}
\end{multline}
\section{Experiments}
\label{section4}
\subsection{Settings}
The MOVI-Codec networks were trained end-to-end on the  Kinetics-600 dataset\cite{kay2017kinetics,carreira2018short} and the Vimeo-90K dataset\cite{xue2019video}. The Kinetics-600 videos are downloaded 
from YouTube, each video having duration of about 10s and various resolutions and frame rates. We used part of the testing set from Kinetics-600, which consists of around 10,000 videos, to conduct our experiments. From each video, a random 128 $\times$ 128 patch with 49 frames was randomly selected for training, and the values of each input video were normalized to [-1,1]. We randomly downsampled the original frame and extracted a 128 $\times$ 128 patch to reduce any previously introduced compression artifacts. The Vimeo-90K dataset consists of 4,278 videos of fixed resolution 448 $\times$ 256. Since the Vimeo-90K dataset has 7 frames per video, we randomly selected a patch of the same size as mentioned before with 7 frames for training. In the Vimeo-90K dataset, the consecutive frames are selected so that the average motion magnitude is between 1-8 pixels, whereas there is no limitation to the motion magnitude between frames in the Kinetics-600 dataset. The mini-batch size is set as 8 for training, and the step length $N$ in our recurrent network is set as 7. By training on both the Vimeo-90K and the Kinetics-600 dataset, we are able to generalize our model to a wide range of natural motions. We tested the MOVI-Codec on the VTL dataset \cite{vtl}, the JCT-VC \cite{bossen2013common} (Class B, C, D and E) datasets, and the UVG datasets \cite{uvg}. These datasets cover a variety of resolutions as shown in Table \ref{dataset}. For fair comparison with \cite{lu2019dvc, lu2020end} and \cite{yang2020learning}, we tested our framework on the JCT-VC datasets using the first 100 frames, and tested on VTL and UVG using all frames.

\begin{table*}[htp]
\centering
\caption{Resolutions of different datasets used for evaluation}
\label{dataset}
\begin{tabular}{|c|c|c|c|c|c|c|}
\hline
        Dataset & VTL  & UVG & JCT-VC Class B & JCT-VC Class C& JCT-VC Class D& JCT-VC Class E   \\ \hline
Resolution &  352 $\times$ 288 & 1920 $\times$ 1080 & 1920 $\times$ 1080& 832 $\times$ 480& 416 $\times$ 240& 1280 $\times$ 720 \\ \hline
\end{tabular}
\end{table*}

To evaluate the quality of the reconstructed videos, we used two quality models: the perception-based MS-SSIM \cite{wang2003multiscale} and the non-perceptual PSNR. Multiscale SSIM (MS-SSIM)  is a widely used image quality assessment model which
captures local luminance, contrast, and structural information. For each quality metric, we trained 5 models with different values of the weighting parameter $\lambda$ to cover different bitrate ranges. For the MS-SSIM model, $\lambda$ was set to 0.01, 0.05, 0.1, 0.5 and 1.0, respectively. For PSNR based models, $\lambda$ was set to 0.0005, 0.0025, 0.005, 0.025 and 0.05. We fixed $\beta = 1$, since we didn't observe any significant differences in model performance as it was varied over the range 0.1 to 10.0.

We compared our method with both traditional and recent deep learning models. H.264 \cite{wiegand2003overview}, H.265 \cite{sullivan2012overview} and the most recent H.266 \cite{bross2021overview} were included as representatives of traditional hybrid compression codecs. We follow \cite{lu2019dvc} \cite{yang2020learning}, and used the x264 and x265 ``LDP very fast'' mode. For H.266, we followed \cite{h266} to implement the  ``faster'' mode. So that we could compare against another motion-free method, we also included the H.265 zero motion setting, using x265 with \textit{merange} set to zero, which allows exploiting temporal redundancy using an IB prediction structure but without performing  motion estimation. In this regard, this setting is most similar to our architecture \cite{brites2015distributed}. Among recent deep learning models, DVC \cite{lu2019dvc} and Wu \textit{et al.} \cite{wu2018video} are optimized for PSNR, Habibian \textit{et al.} \cite{habibian2019video} and Cheng \textit{et al.} are optimized for MS-SSIM, and HLVC \cite{yang2020learning} has both MS-SSIM optimized and PSNR optimized results. 

\subsection{Results}
\begin{figure*} [!t]
\centerline{
   \label{visual}
\includegraphics[width=2\columnwidth]{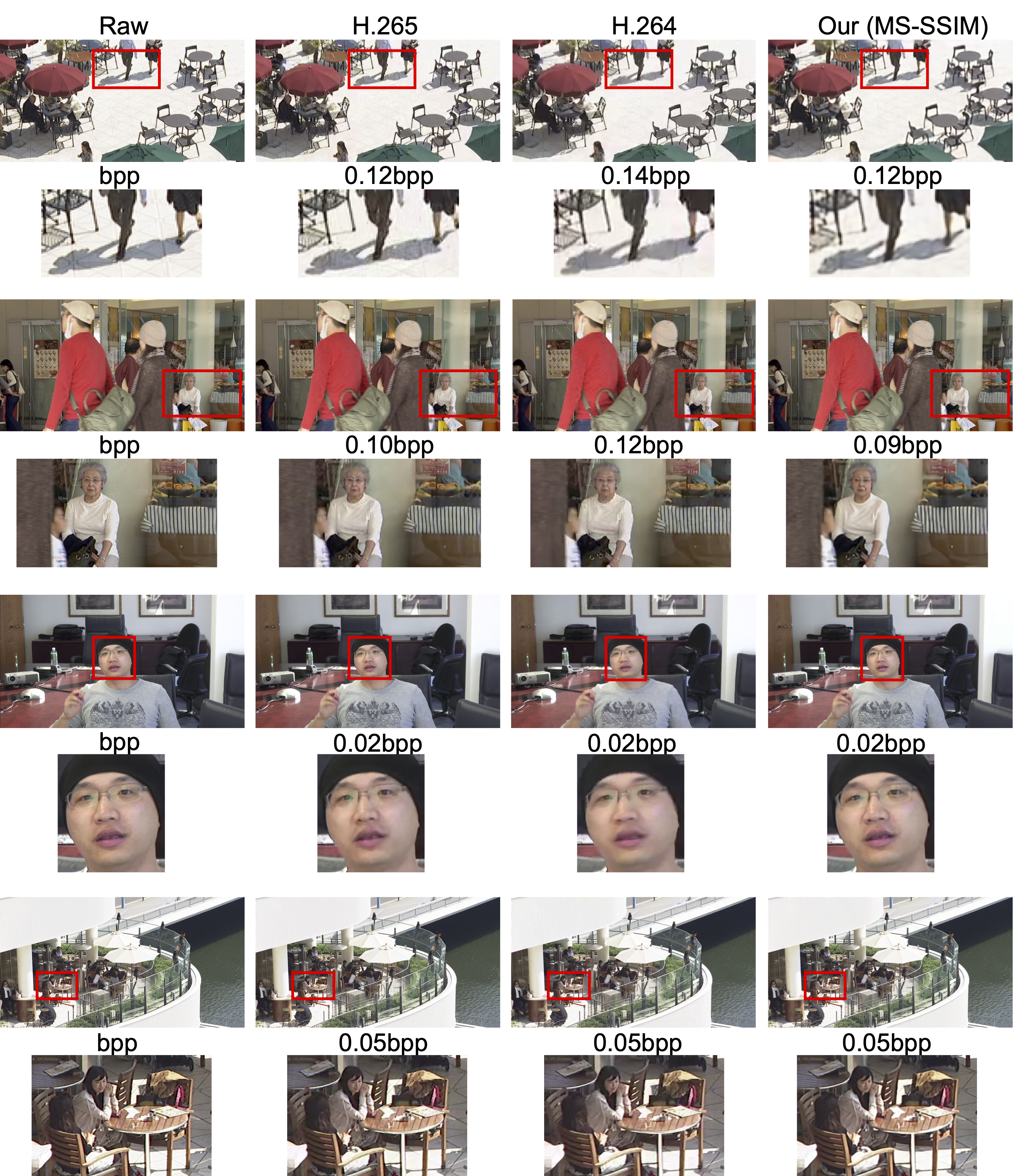}}
\caption{Visual examples of our method as compared with H.264 and HEVC. }
\label{fig:visual}
\end{figure*}

In this section, we compare our video compression engine against the standards H.264, HEVC, and H.266/VVC, and with other deep learning-based video compression architectures (Wu \cite{wu2018video}, DVC \cite{lu2019dvc, lu2020end}, and Cheng \cite{cheng2019learning}) on the  UVG  dataset,  the  VTL dataset,  and  the  HEVC  Standard  Test  Sequences  (Class  B, Class C, Class D, and Class E). When compressing videos using the H.264 and HEVC codecs, we followed the settings in \cite{lu2019dvc} and used FFmpeg with the \textit{very fast} mode\footnote{
H.264: \textit{ffmpeg -pix\_fmt yuv420p -s WxH -r FR -i Video.yuv -vframes N -c:v libx264 -preset veryfast -tune zerolatency -crf Q -g GOP -bf 2 -b\_strategy 0 -sc\_threshold 0 output.mkv}

H.265: \textit{ffmpeg -pix\_fmt yuv420p -s WxH -r FR -i Video.yuv -vframes N -c:v libx265 -preset veryfast -tune zerolatency -x265-params ``crf=Q:keyint=GOP'' output.mkv}

\textit{FR, N, Q, GOP} represents the frame rate, the number of encoded frames, quality, GOP size, respectively. \textit{N} is set to 100 for HEVC datasets.}. When implementing H.266, we followed \cite{h266} using the \textit{faster} mode. We also provide visual examples of our approach against other approaches in Figure \ref{fig:visual}. More exemplar reconstructed videos are included on our project page with link given in the Abstract.

\begin{figure*} [!ht]
\centerline{
   \label{MS-SSIM}
\includegraphics[width=0.63\columnwidth]{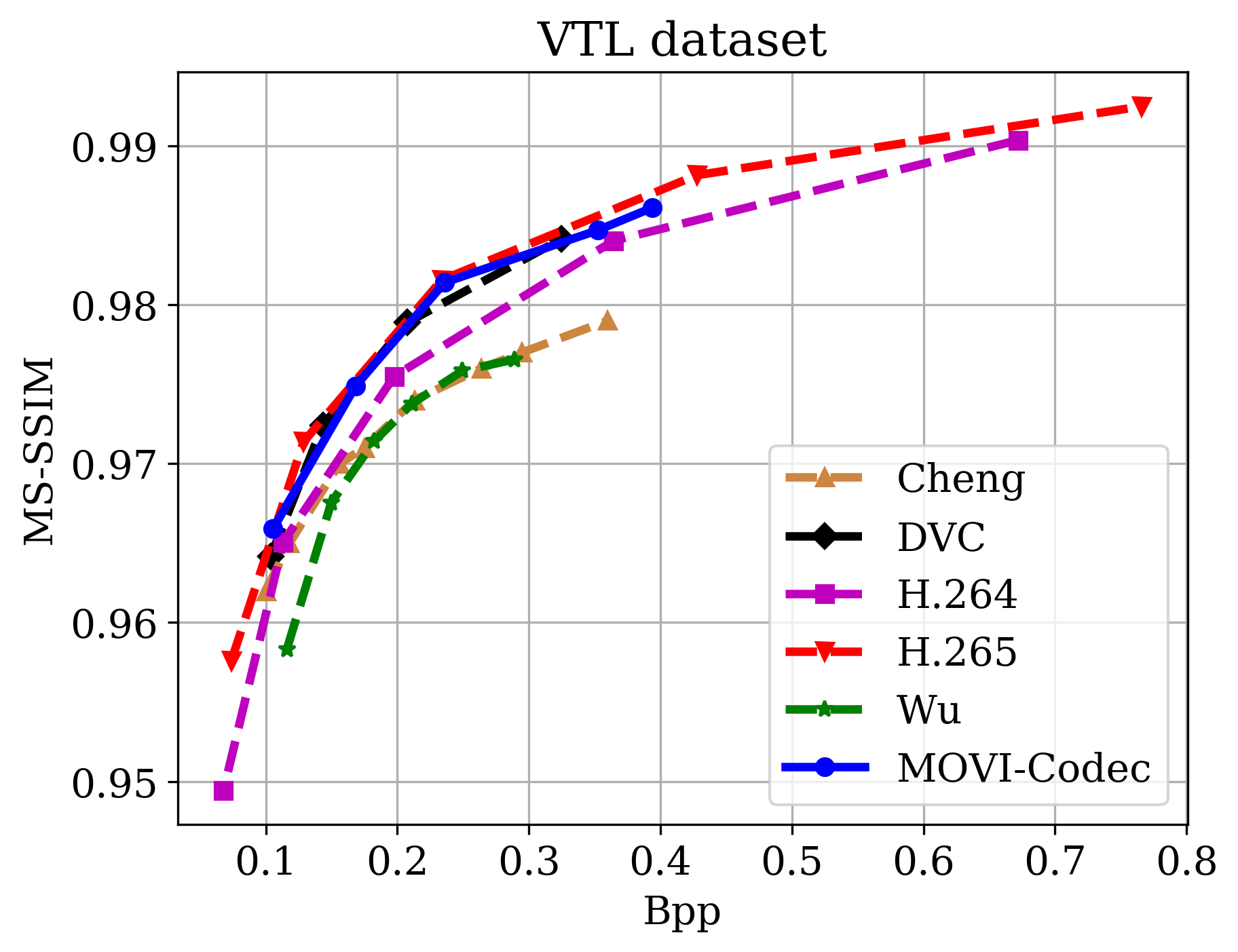}}
\caption{MS-SSIM on the VTL dataset (352 $\times$ 288) for different compression codecs. Our method is  competitive with the state of the art over varying bit rates on these low-resolution videos.}
\label{fig:VTL}
\end{figure*}

\begin{figure*} [!ht]
\centerline{
\subfigure{
  \label{PSNR}
\includegraphics[width=0.7\columnwidth]{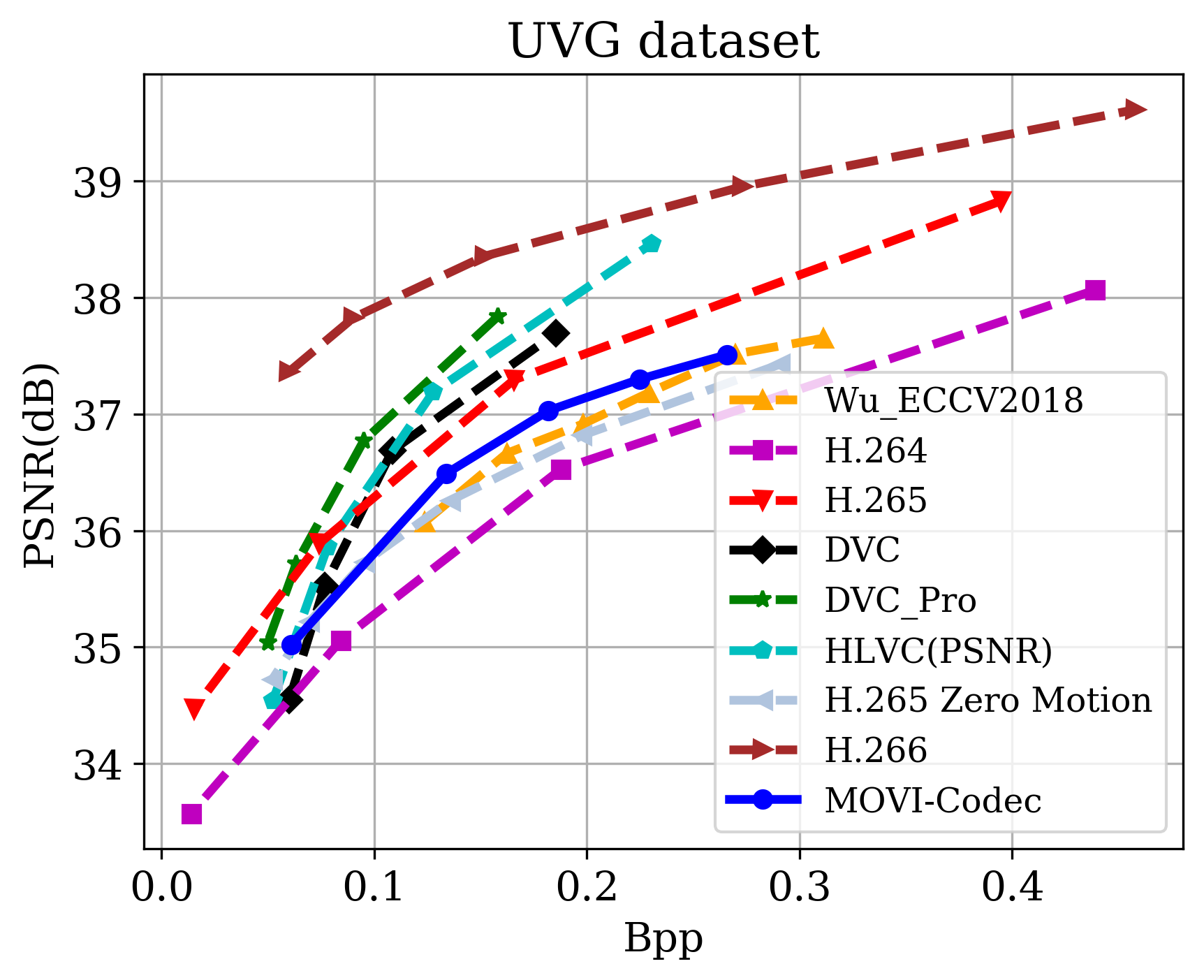}}
\subfigure{
  \label{MS-SSIM}
\includegraphics[width=0.722\columnwidth]{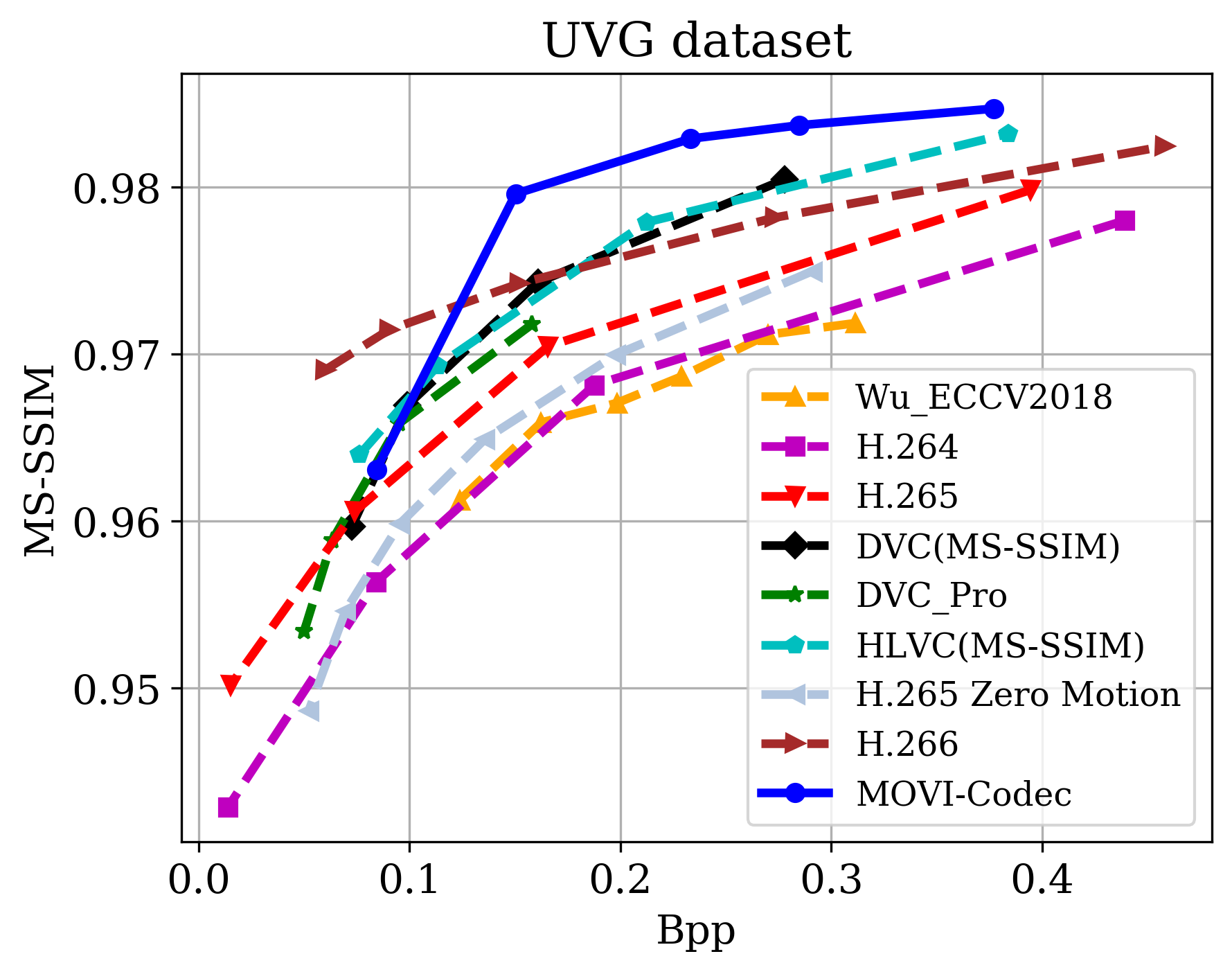}}

}

\caption{PSNR and MS-SSIM on the UVG dataset (1920 $\times$ 1080) for different compression codecs. Our method outperformed all compression methods against the perceptually relevant MS-SSIM, while remaining highly competitive against the non-perceptual PSNR. }
\label{fig:UVG}
\end{figure*}

\begin{figure*} [!ht]
\centerline{
\subfigure{
  \label{MS-SSIM}
\includegraphics[width=0.7\columnwidth]{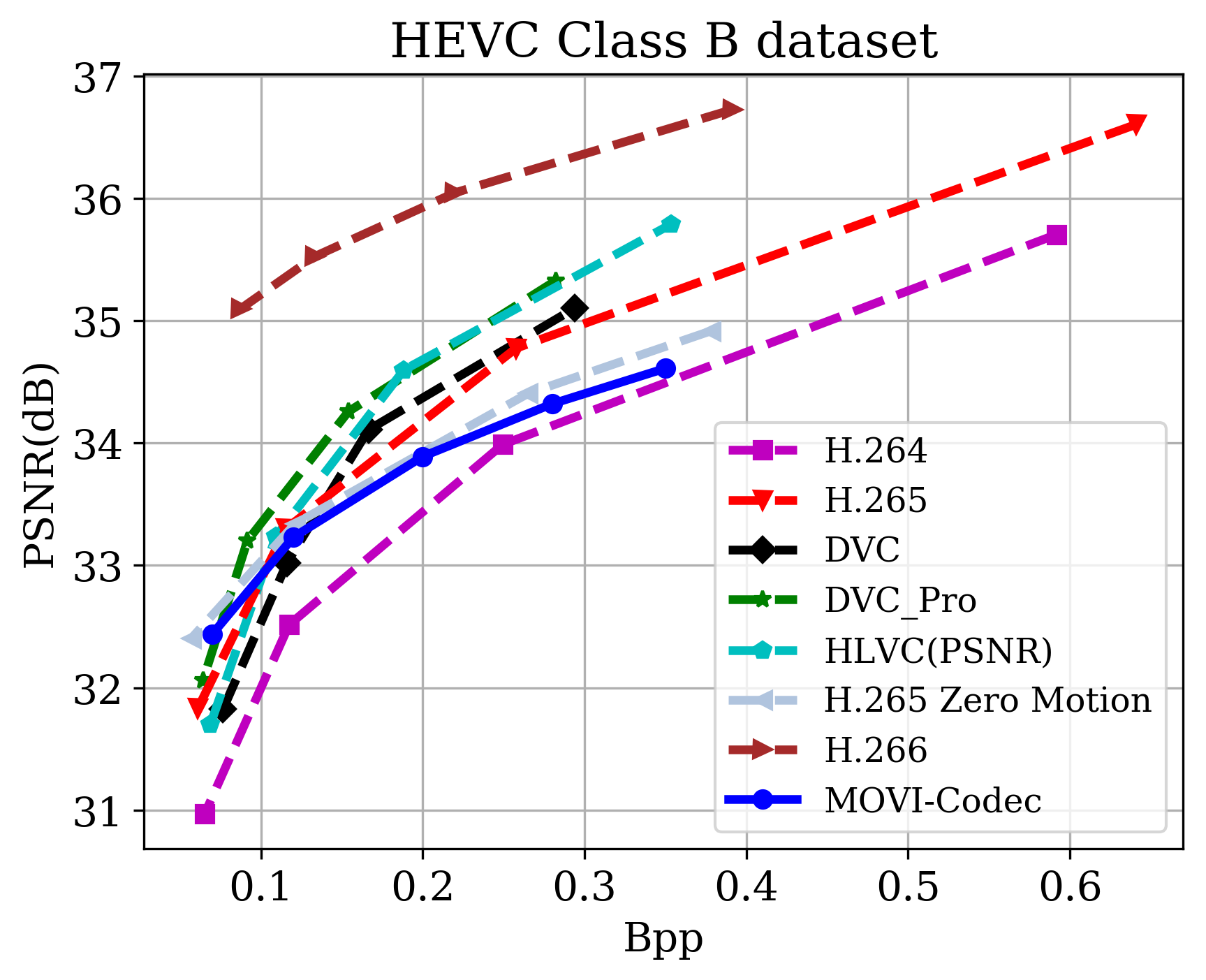}} 
\subfigure{
  \label{PSNR}
\includegraphics[width=0.7\columnwidth]{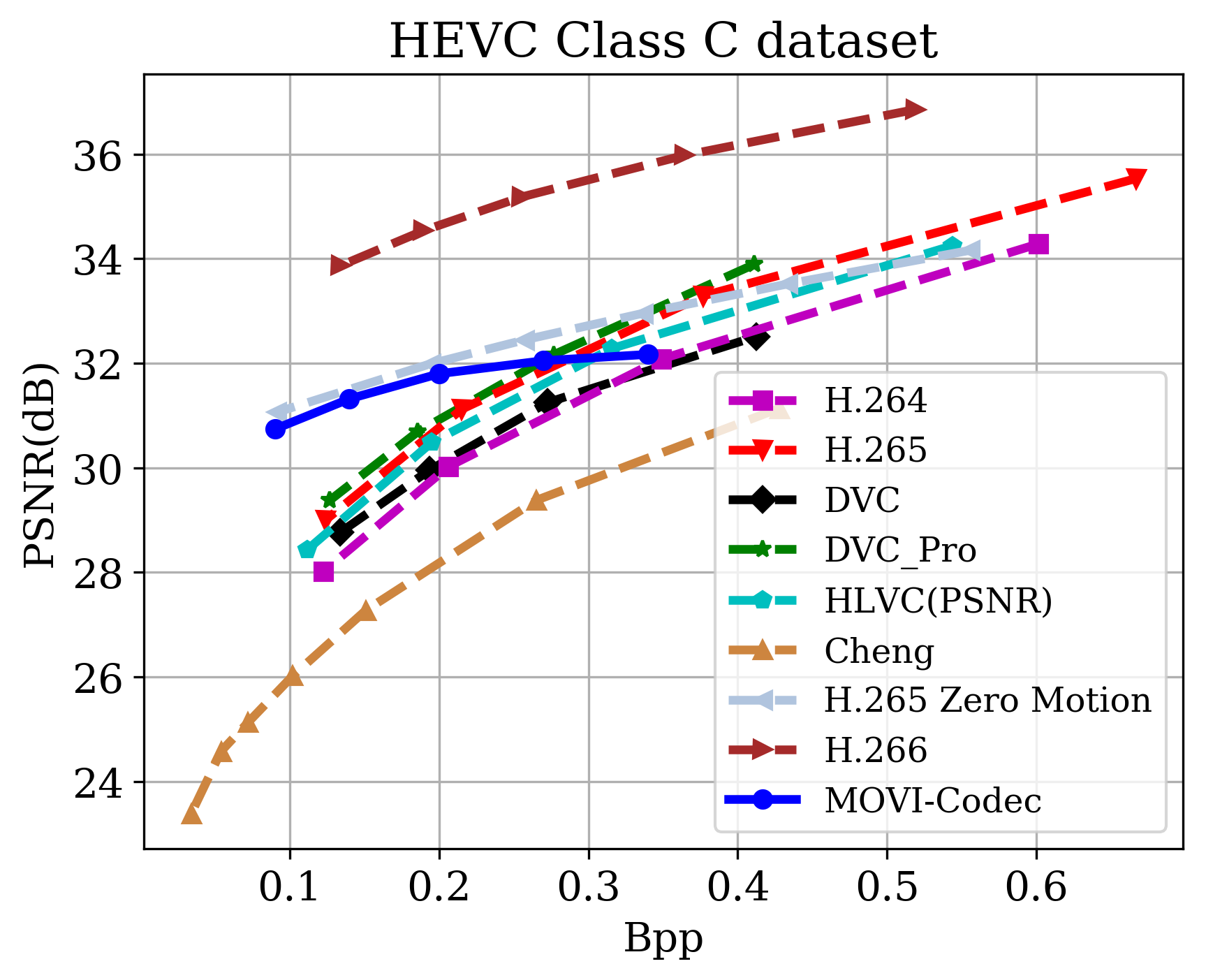}}
}
\centerline{
\subfigure{
  \label{MS-SSIM}
\includegraphics[width=0.7\columnwidth]{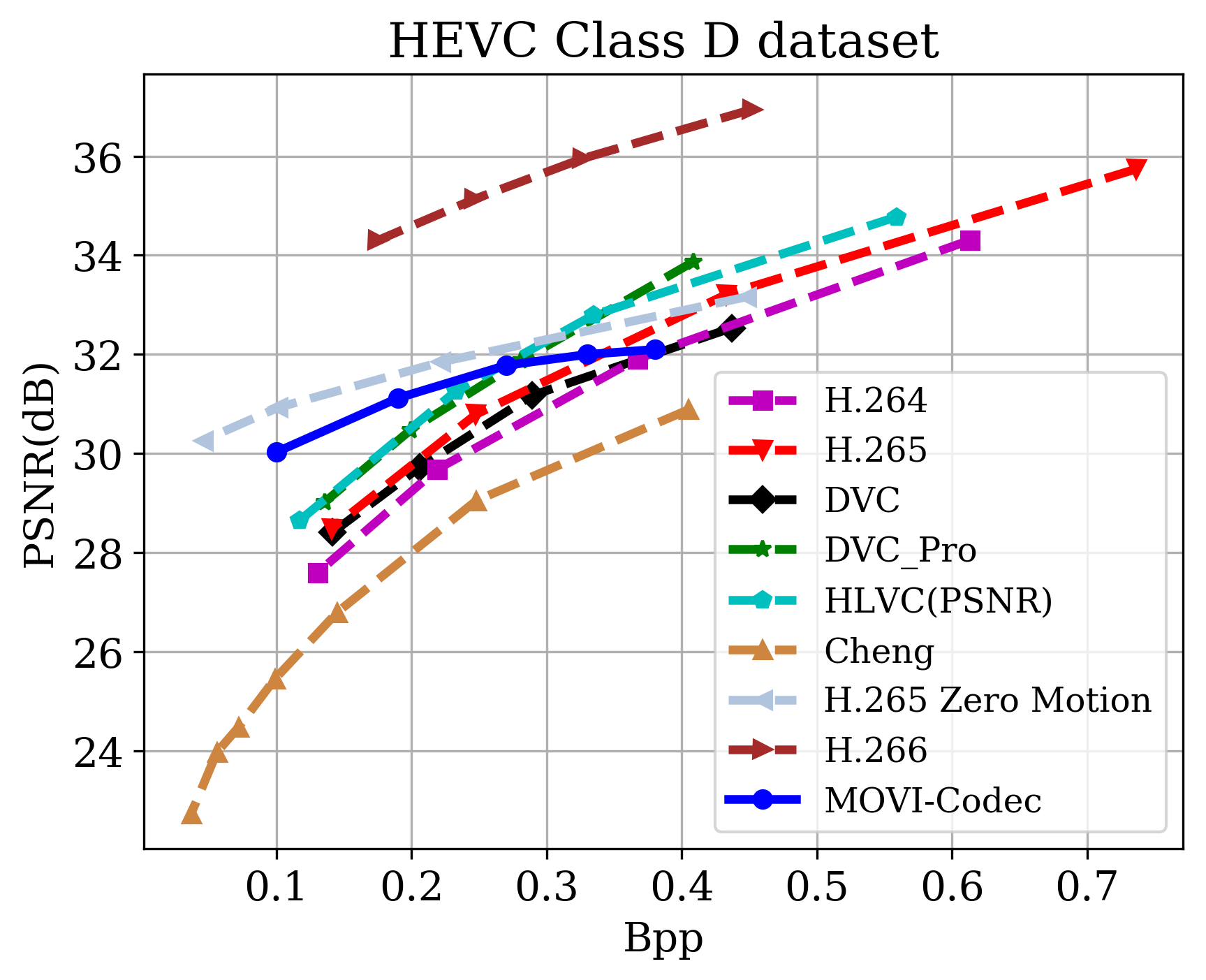}} 
\subfigure{
  \label{PSNR}
\includegraphics[width=0.7\columnwidth]{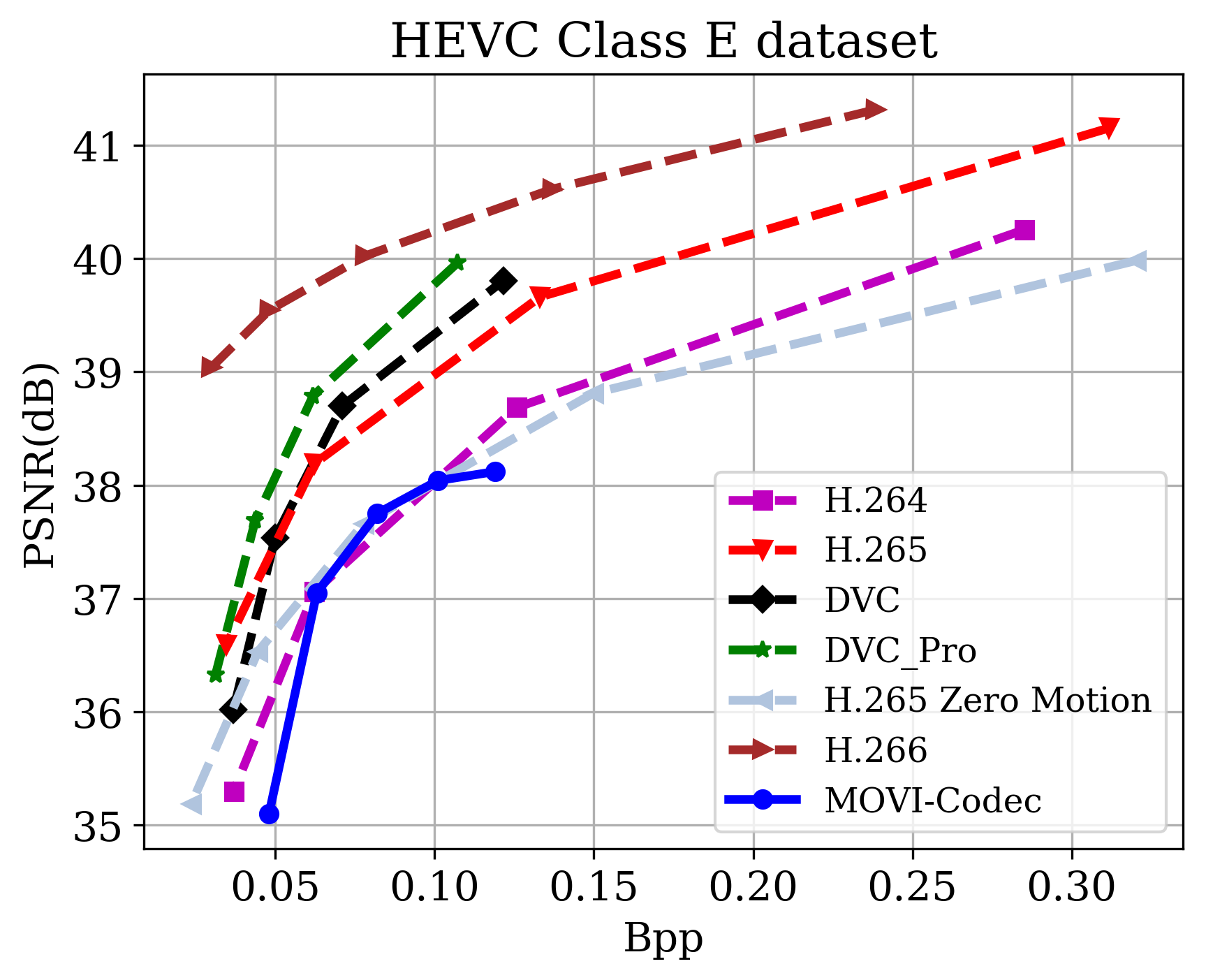}}
}
\caption{PSNR of HEVC test sequences for different compression codecs. The resolution of Class B is 1920 $\times$ 1080, of Class C is 832 $\times$ 480, of Class D is 416 $\times$ 240 and of Class E is 1280 $\times$ 720. Overall, our method is competitive with H.265, and is particularity good at lower bit rates on lower resolution datasets.}
\label{fig:HEVC_psnr}
\end{figure*}

\begin{figure*} [!ht]
\centerline{
\subfigure{
  \label{MS-SSIM}
\includegraphics[width=0.7\columnwidth]{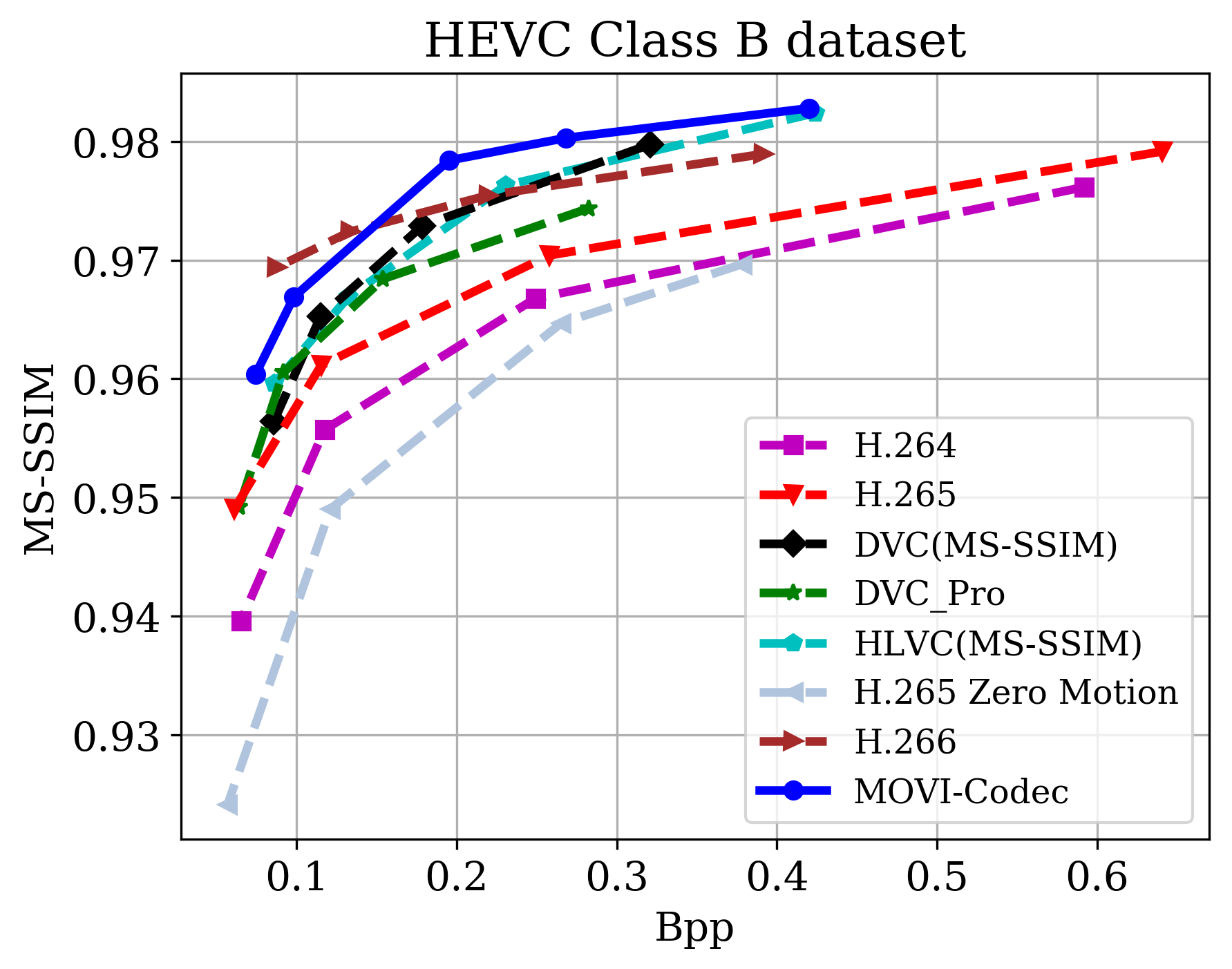}} 
\subfigure{
  \label{PSNR}
\includegraphics[width=0.7\columnwidth]{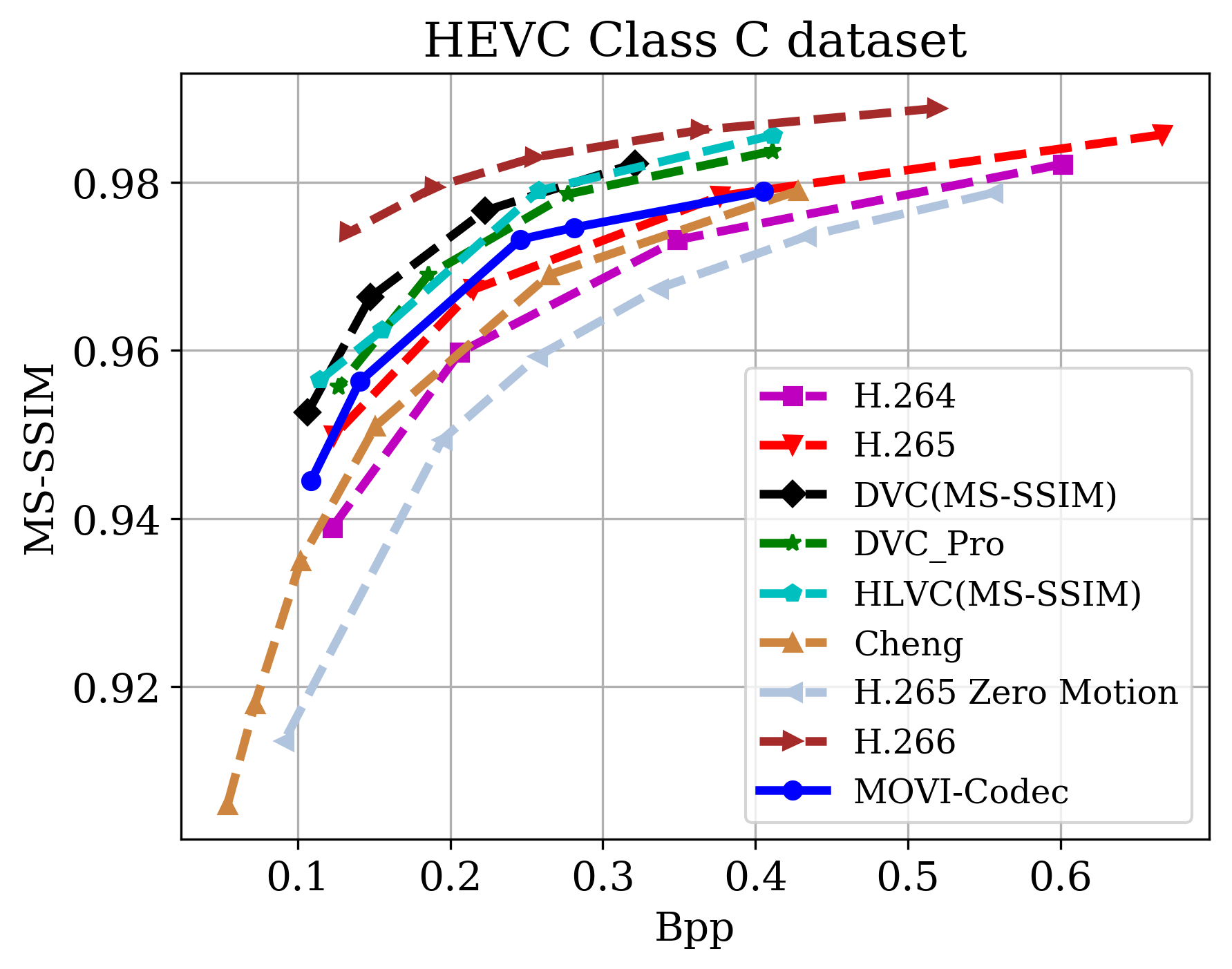}}
}
\centerline{
\subfigure{
  \label{MS-SSIM}
\includegraphics[width=0.7\columnwidth]{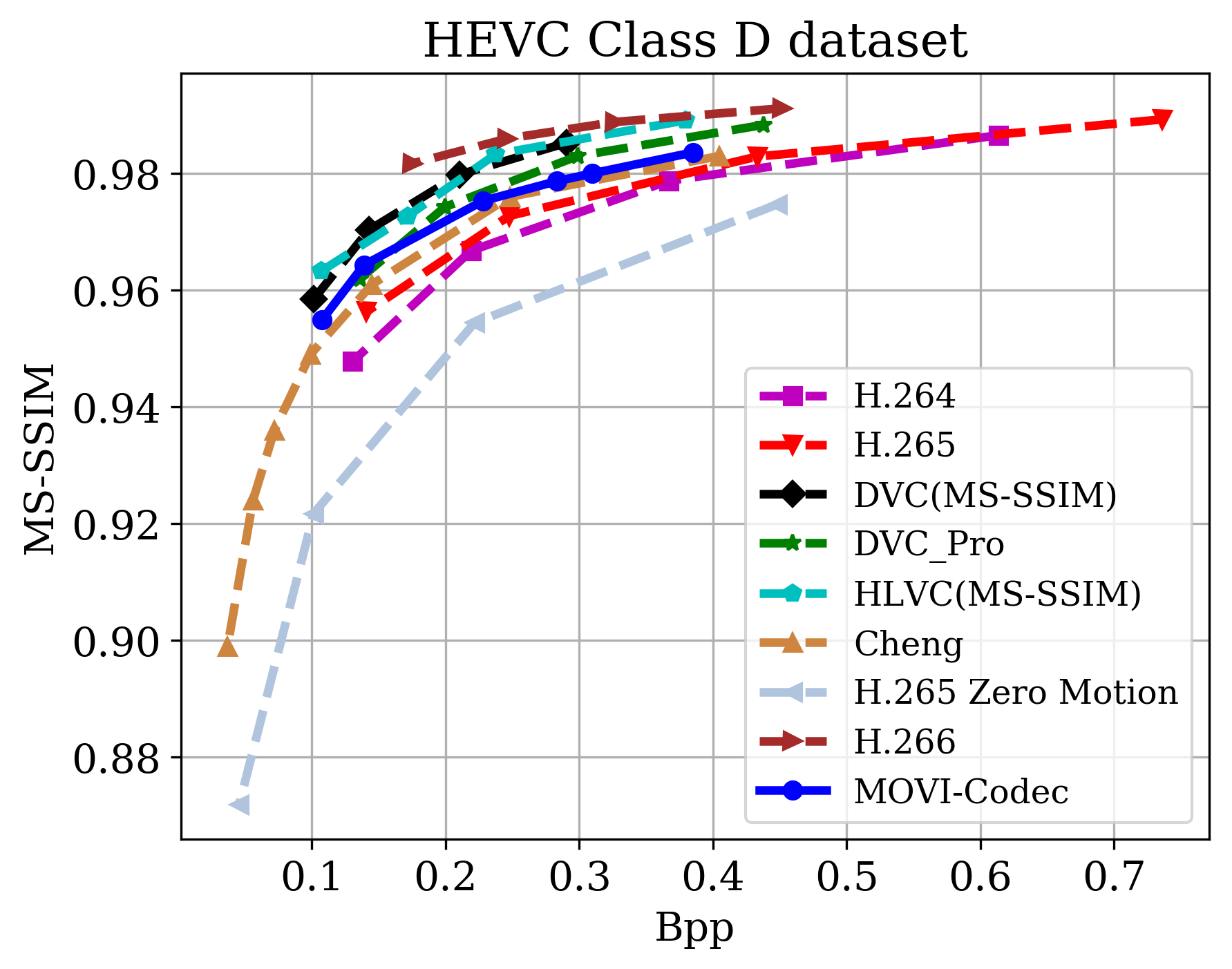}} 
\subfigure{
  \label{PSNR}
\includegraphics[width=0.7\columnwidth]{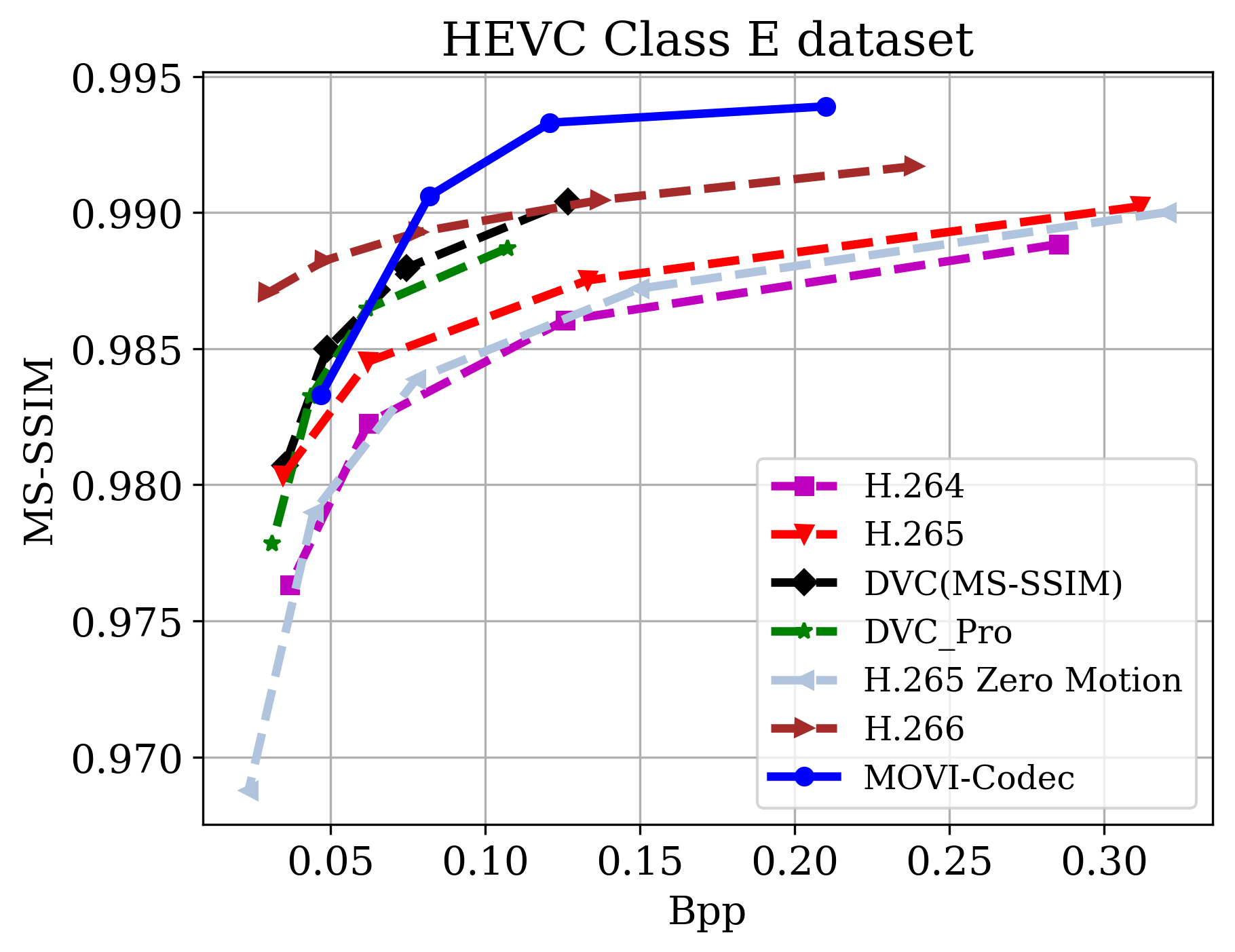}}
}
\caption{MS-SSIM of HEVC test sequences for different compression codecs, where the resolution of Class B is 1920 $\times$ 1080, of Class C is 832 $\times$ 480, of Class D is 416 $\times$ 240 and of Class E is 1280 $\times$ 720. Our method outperformed H.265 and is competitive with other state of the art deep learning models.}
\label{fig:HEVC}
\end{figure*}

% \subsubsection{Rate-distortion Curve}
Figures \ref{fig:VTL}, \ref{fig:UVG}, \ref{fig:HEVC_psnr}, and \ref{fig:HEVC} show the experimental results on the  VTL  dataset,  the  UVG dataset,  and  the  HEVC  Standard  Test  Sequences  (Class  B, Class C, Class D, and Class E). These results show that our network outperformed both H.264 and the  HEVC standard against MS-SSIM. On datasets with higher resolution videos (UVG dataset, HEVC Class B dataset, and HEVC Class E dataset), our network was able to outperform the latest H.266 codec at higher bitrates as assessed using the perceptually relevant MS-SSIM algorithm. We also compared our model against several deep learning-based compression models, including a frame interpolation-based model by Wu \textit{et al.}\cite{wu2018video}, DVC \cite{lu2020end}, HLVC \cite{yang2020learning} and the video compression framework proposed by Cheng \textit{et al.}, which uses an added spatial energy compaction penalty in the loss function \cite{cheng2019learning}. Among these, DVC and HLVC were trained  on both PSNR and MS-SSIM, to obtain better results against each metric. In our comparison, we include the best performance for these two methods for each metric. It is worth noting that our model only uses one previous frame as input, whereas in Wu's framework, both neighboring frames are utilized when reconstructing the middle frame. Additionally, our framework replaces the classical motion estimation and compensation module by instead training the network to optimally interpolate  displaced frame differences. For completeness, we also evaluated all models against the PSNR, where MOVI-Codec did not always perform as well. However, this is a problem with the PSNR, which is not perceptually relevant, and which produces significantly inferior quality predictions than perception-based quality predictors like MS-SSIM \cite{wang2009mean}. Indeed, the high quality of the reconstructions that we make available on the model page (see link in Abstract) further attests to this. As has been observed by others \cite{balle2019end,rippel2017real,rippel2019learned, agustsson2019generative}, perceptual measures are better arbiters of deep compressed video quality than absolute fidelity models like the PSNR. It is worth noting that when comparing our model against the H.265 zero motion setting, while both methods do not utilize motion estimation, MOVI-Codec was able to perform better with respect to MS-SSIM than the H.265 zero motion setting, while delivering similar performance against PSNR.

% \subsubsection{Bit-rate Reduction}
% We also evaluate our method quantitively using Bj{\o}ntegaard Delta Bit-Rate (BDBR) and BD-PSNR (BD-MSSSIM) \cite{bjontegaard2001calculation} with the anchor of H.265. In the video compression task, BDBR and BD-PSNR (BD-MSSSIM) are widely used to evaluate the performance of different video compression systems. BDBR represents the average percentage of bit rate savings when compared with the baseline algorithm at the same PSNR (MS-SSIM). BD-PSNR (BD-MSSSIM) represents the gain (dB) when compared with the baseline algorithm with same bit rate. \todo{BDBR and BD-PSNR}
% % \input{BDBR_table}

\subsection{Ablation Studies}
We conducted ablation studies to assess the choices we made in our approach, specifically with respect to the choice of displaced frame differences, and the effectiveness of the proposed LSTM-UNet. The results are shown in Figure \ref{fig:ablation_dis} and Figure \ref{fig:ablation_lstm}.

\subsubsection{Displaced Frame Difference Combination}
\begin{figure} [!t]
\centerline{
   \label{ablation_dis}
\includegraphics[width=0.7\columnwidth]{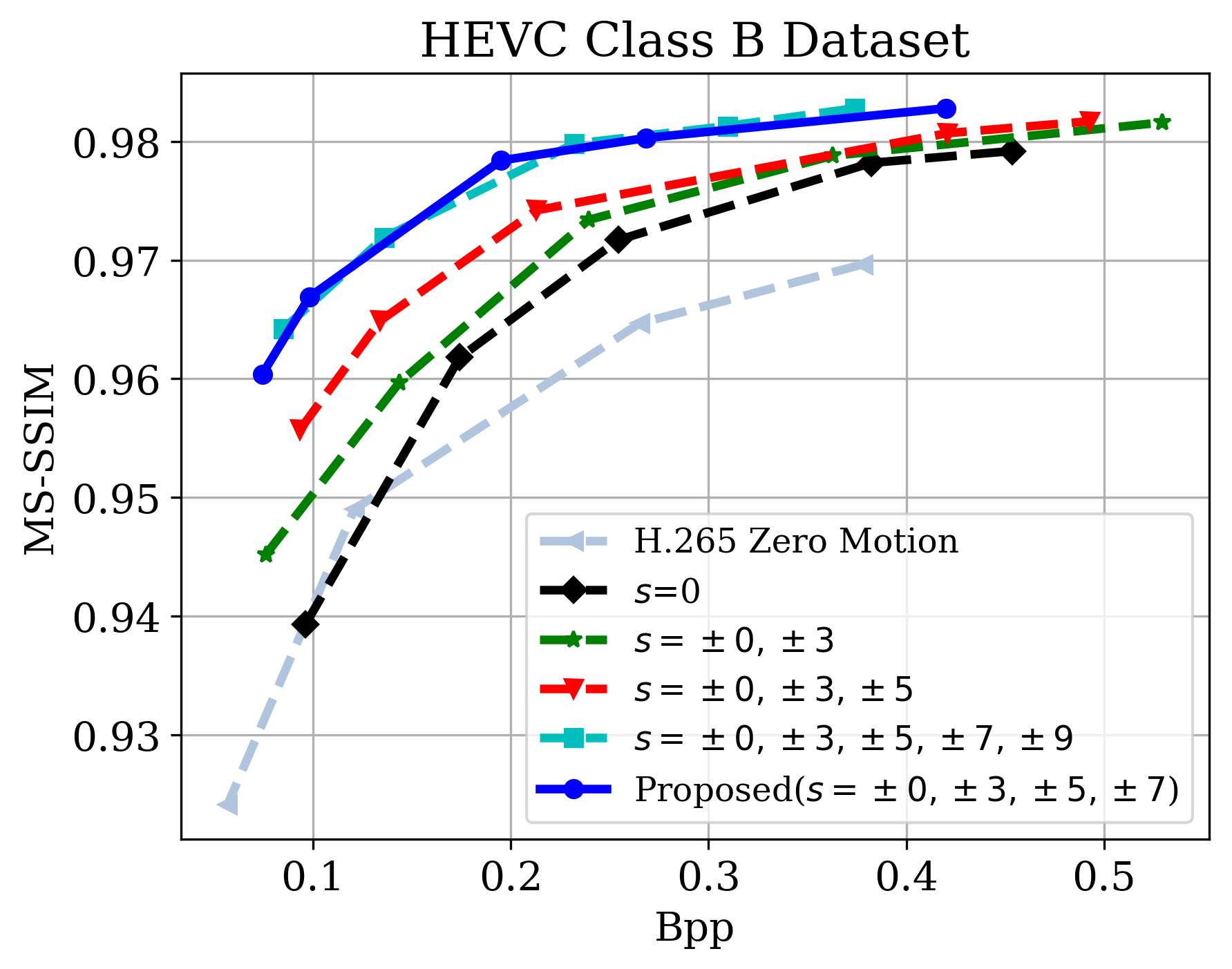}}
\caption{Ablation study of displaced frame difference combinations. }
\label{fig:ablation_dis}
\end{figure}

Figure \ref{fig:ablation_dis} shows the experimental results on different combinations of displaced frame differences, where $s = 0$ refers to frame differences with no displacements, which gives the worst performance of all combinations evaluated. This shows the value of  ``displaced'' frame differences as a way of training the network on more diverse motion induced displacements. Including displacements as large as $s = 7$ greatly increases the overall performance, by allowing interpolation of larger motions in videos. We also tried adding $s = 9$ to our choice of displacement combinations, but this new combination did not improve the overall performance, meaning that our combination of displacements was adequate to capture motions of various sizes. It is worth noting that as compared with the H.265 zero motion configuration, which also does not utilize motion estimation, our network was able to perform better as assessed by the perceptually relevant MS-SSIM, including when $s = 0$.

\subsubsection{Effectiveness of the LSTM-UNet}
\begin{figure} [!t]
\centerline{
   \label{ablation_lstm}
\includegraphics[width=0.7\columnwidth]{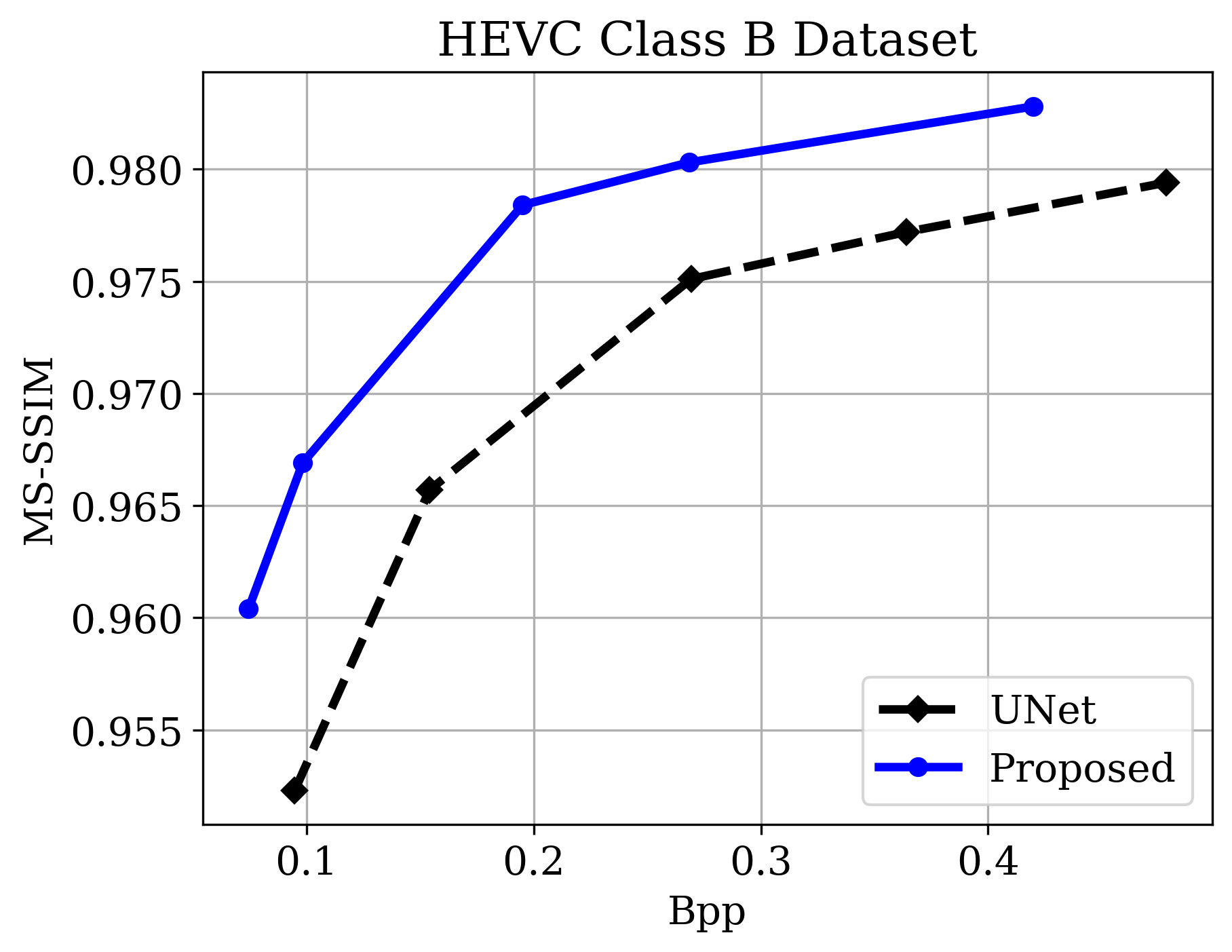}}
\caption{Ablation study of the effectiveness of the proposed LSTM-UNet. }
\label{fig:ablation_lstm}
\end{figure}

Figure \ref{fig:ablation_lstm} shows the experimental results on the HEVC Class B dataset when using UNet and LSTM-UNet to reconstruct frames, respectively. As shown in the example, LSTM-UNet extends the advantage of UNet for extracting and representing spatial descriptors to include spatio-temporal descriptors using C-LSTM blocks, yielding better reconstruction performance. In addition, LSTM-UNet converges faster than the UNet counterparts, shortening the training time of the network. 

\subsection{Motion Vector Analysis}
\begin{figure*} [!ht]
\centerline{
\subfigure{
  \label{PSNR}
\includegraphics[width=0.7\columnwidth]{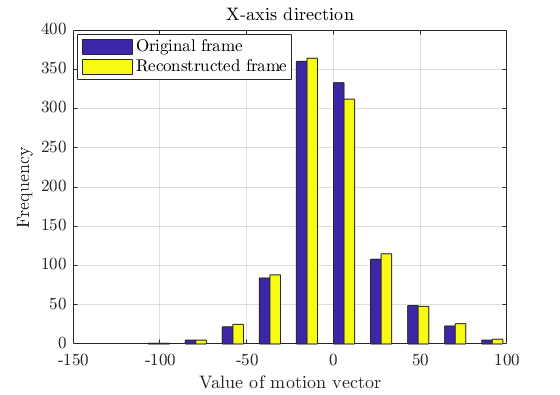}}
\subfigure{
  \label{MS-SSIM}
\includegraphics[width=0.7\columnwidth]{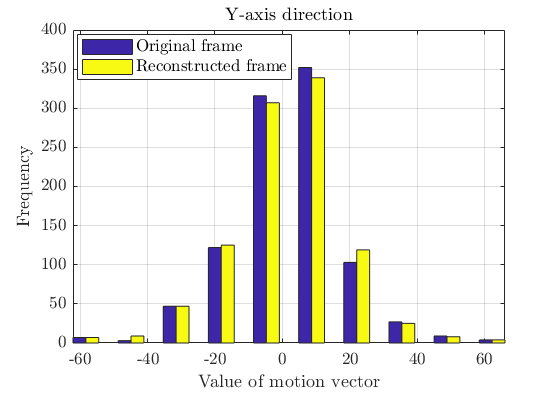}}
}
\caption{Distributions of the maximum and minimum motion vector components along the horizontal (left) and vertical (right) axes of the HEVC Class B dataset.}
\label{fig:MV_axis}
\end{figure*}

To verify that MOVI-Codec can capture large motions with the chosen set of displacement combination, we calculated the optical flow of adjacent frames in the testing datasets using a pre-trained network called SPynet\cite{ranjan2017optical}. To emphasize large motions, we calculated all motion vectors against adjacent frames, and only picked the minimum and maximum motion vectors in the x and y directions. As a result, we ended up with four values of motion vectors for each adjacent frames. Figure \ref{fig:MV_axis} shows the distribution of the picked motion vectors on all videos in the HEVC Class B dataset, which is the dataset having the highest resolution videos among our testing datasets. From the figure, we can conclude that our model produced a similar distribution as the original frame pairs, hence our model was able to capture large motions using a set of small displacements.

\begin{figure*} [!ht]
\centerline{
\subfigure[Original frame]{
  \label{PSNR}
\includegraphics[width=0.63\columnwidth]{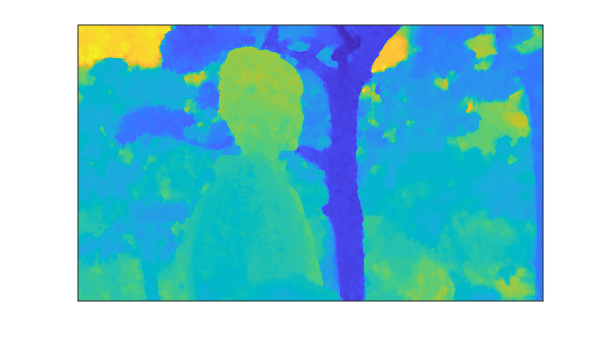}}
\subfigure[Reconstructed frame]{
  \label{MS-SSIM}
\includegraphics[width=0.7\columnwidth]{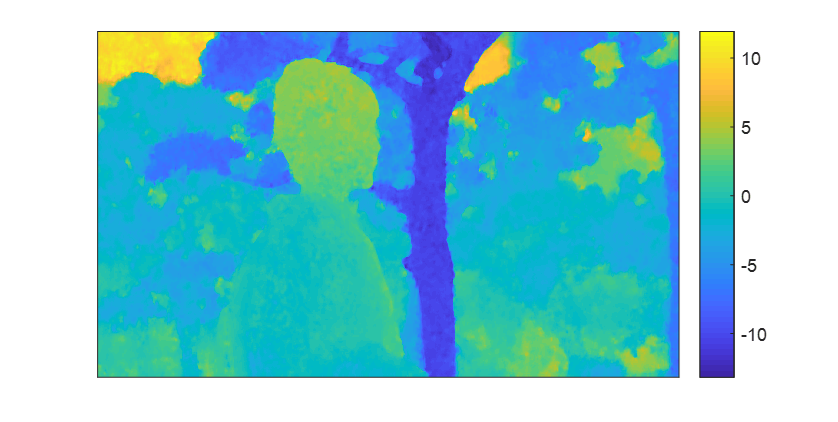}}}
\caption{Optical flow along the horizontal direction between two adjacent frames in the Kimono video.}
\label{fig:kimono_MV}
\end{figure*}

\begin{figure*} [!ht]
\centerline{
\subfigure[Original frame]{
  \label{PSNR}
\includegraphics[width=0.63\columnwidth]{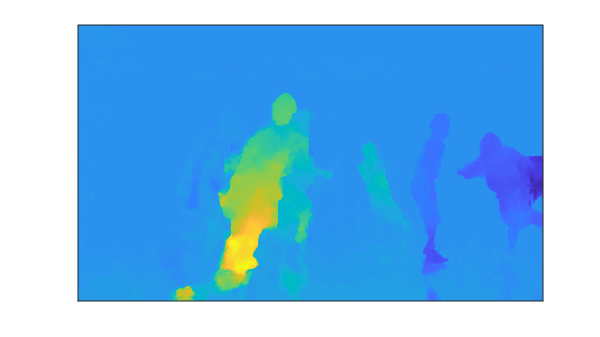}}
\subfigure[Reconstructed frame]{
  \label{MS-SSIM}
\includegraphics[width=0.7\columnwidth]{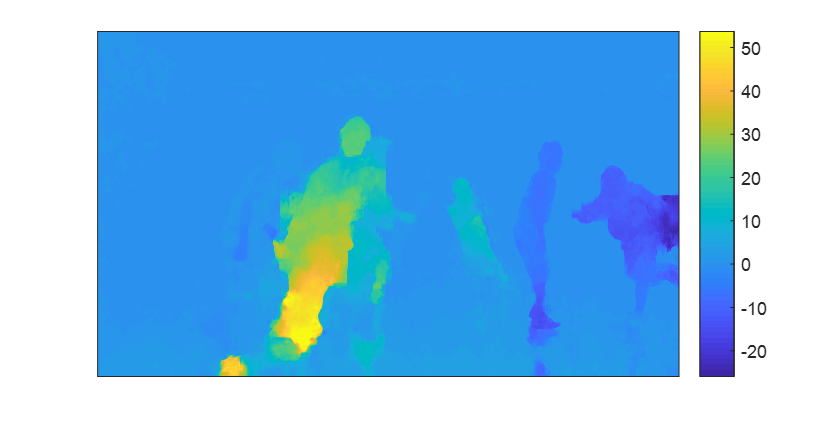}}
}
\caption{Optical flow along the horizontal directionbetween two adjacent frames in the Basketball Drive video.}
\label{fig:basketball_MV}
\end{figure*}

Figure \ref{fig:kimono_MV}  and \ref{fig:basketball_MV} illustrate the accuracy of our motion reconstruction. The test video in Figure \ref{fig:kimono_MV} shows the x axis optical flow between two adjacent frames from the Kimono video, which is a video with a moving background and slow motion, whereas Figure \ref{fig:basketball_MV} shows the optical flow images of two adjacent frames in Basketball Drive video, which has a static background and large motions. In both videos, our model was able to reconstruct motion accurately. 

\subsection{Model Analysis}
\begin{figure} [!t]
\centerline{
\includegraphics[width=0.6\columnwidth]{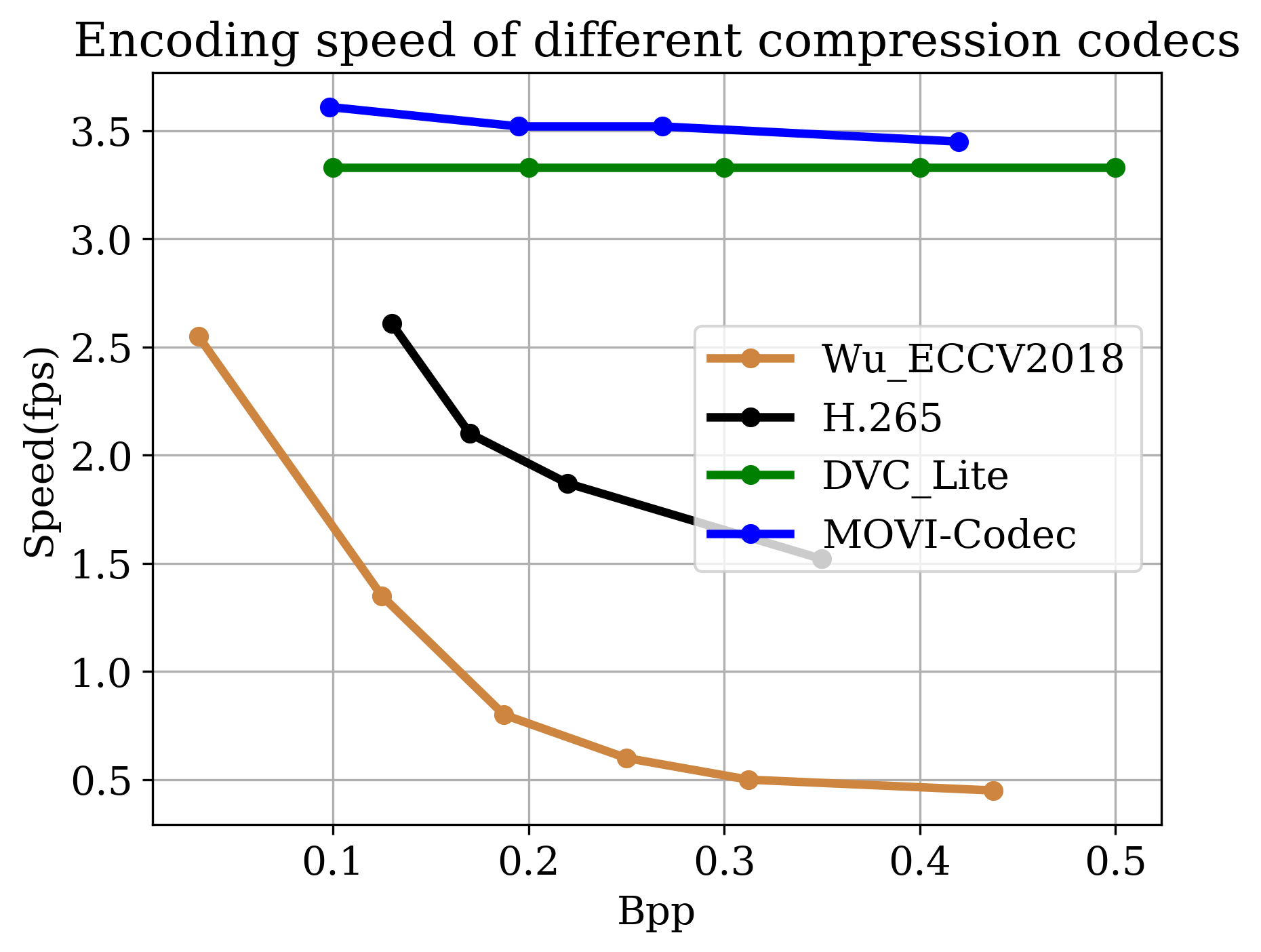}}
\caption{Encoding speed of different compression codecs. H.265 refers to the encoding speed of the x265 codec \textit{slower} setting. }
\label{fig:speed}
\end{figure}

To compare the computational complexity of the different  codecs, we tested two deep learning models: the one proposed by Wu \textit{et al.} \cite{wu2018video}, the light version of DVC called DVC Lite \cite{lu2020end}, and the commercial software x265 for H.265 compression, using a server with an Intel Core i9-9940X CPU and GTX 1080Ti on video sequences of resolution 1920 $\times$ 1080. The experimental results are provided in Figure \ref{fig:speed}. 

The overall encoding speed of our framework is mostly invariant of bitrate, whereas since Wu's framework adopts a progressive coding scheme, its encoding speed varies with the target bitrate. In our framework, although we adopted an RNN-based compression method on displaced frame differences, we utilized the RNN unit to store temporal dependencies and did not use a progressive coding scheme for compression. DVC Lite is a lightweight version of DVC with a more efficient motion estimation module and a lightweight motion compression network, which can be twice as fast as the original DVC model in terms of encoding speed\cite{lu2020end}. Our framework is faster than the lightweight model, further justifying the use of learned interpolation of displaced frame differences. Since the arithmetic coding at lower bitrates is faster than at larger ones, there is a slight slope to our encoding speed curve. But overall, the complexity of our model is invariant to bitrate, which means that our model maintains a stable encoding speed regardless of video content or bitrate for a given resolution.

As shown in Figure \ref{fig:speed}, compared with the traditional hybrid codec, our model is faster than the latest codec HEVC with \textit{slower} setting. However, using the \textit{very fast} setting on x264 and x265, the encoding speed can run at 110 fps and 30 fps, respectively. Of course, by applying model acceleration techniques such as model distillation, model quantization, or by decreasing the model size, it should be possible to similarly accelerate the encoding speed of our framework.
\subsection{Discussion}
From Figures \ref{fig:VTL}, \ref{fig:UVG}, \ref{fig:HEVC_psnr}, \ref{fig:HEVC}, and \ref{fig:speed}, we can conclude that our model delivers better compression performance than LDP \textit{veryfast} setting of traditional hybrid codecs like H.264 and HEVC in terms of MS-SSIM, at a low computational complexity. This justifies our use of displaced frame differences as motion information for video compression. Although our model was able to acheive competitive performances as lower settings of traditional codec having low computational complexity, and without the complicated motion estimation and compensation modules other deep learning-based models use, our model did not outperform all of the state-of-the-art models. Nonetheless, the performance achieved by our model provides a new way of motion computation that may prove quite useful for video compression. In our model, we designed the set of spatial displacements used by our network to cover a reasonable range of natural motions. A promising future direction is to automatically assign displacement combinations as a function of resolution. The encoding speed of our model is state-of-the-art among deep learning models, but has not yet been optimized to match compute-optimized traditional codecs like HEVC or VVC, e.g. by model acceleration methods.
\section{Conclusion and Future Work}
In this paper, we proposed an end-to-end deep learning video compression framework that renovated motion prediction. To be specific, we proposed the use of displaced frame differences as indicators of motion information, and fed them into a deep space-time compression network, which learns optimal between-frame interpolated representations to achieve efficiency. Additionally, we proposed a new version of UNet, called LSTM-UNet, that utilizes both spatial and temporal information to conduct frame reconstruction. Our experimental results show that our approach outperforms the LDP \textit{veryfast} setting of the standard codecs H.264 and H.265 in terms of MS-SSIM. In addition, our  network  was able to outperform  the  latest  H.266 codec at higher bitrates as assessed by the perceptual MS-SSIM algorithm, on high resolution videos. The reduced complexity of the framework and the avoidance of motion search could make it easier to implement on resrouce-limited devices, such as smartphones, VR headsets, and AR glasses.
\label{section6} 
\appendices

% % use section* for acknowledgment
\section*{Acknowledgment}
This work is supported by a grant by Facebook and by grant number 2019844 for the National Science Foundation AI Institute for Foundations of Machine Learning (IFML). The authors would like to thank Fabian Mentzer and Ren Yang for insightful discussions about this work.


\ifCLASSOPTIONcaptionsoff
  \newpage
\fi

% \AtEndEnvironment{thebibliography}{
% % all your extra bibitems go here
% \bibitem{pwcreport}
% ``{Perspectives from the Global
% Entertainment & Media Outlook, 2018-2022},"
% \url{https://www.pwc.com/gx/en/entertainment-media/outlook/perspectives-from-the-global-entertainment-and-media-outlook-2018-2022.pdf}, 2018.
% }
% % 

\bibliographystyle{IEEEtran}
\bibliography{IEEEexample}{}

\begin{thebibliography}{10}
\providecommand{\url}[1]{#1}
\csname url@samestyle\endcsname
\providecommand{\newblock}{\relax}
\providecommand{\bibinfo}[2]{#2}
\providecommand{\BIBentrySTDinterwordspacing}{\spaceskip=0pt\relax}
\providecommand{\BIBentryALTinterwordstretchfactor}{4}
\providecommand{\BIBentryALTinterwordspacing}{\spaceskip=\fontdimen2\font plus
\BIBentryALTinterwordstretchfactor\fontdimen3\font minus
  \fontdimen4\font\relax}
\providecommand{\BIBforeignlanguage}[2]{{%
\expandafter\ifx\csname l@#1\endcsname\relax
\typeout{** WARNING: IEEEtran.bst: No hyphenation pattern has been}%
\typeout{** loaded for the language `#1'. Using the pattern for}%
\typeout{** the default language instead.}%
\else
\language=\csname l@#1\endcsname
\fi
#2}}
\providecommand{\BIBdecl}{\relax}
\BIBdecl

\bibitem{index2017cisco}
C.~V.~N. Index, ``Cisco visual networking index: Forecast and methodology,
  2016--2021,'' \emph{Complete Visual Networking Index (VNI) Forecast},
  vol.~12, no.~1, pp. 749--759, 2017.

\bibitem{cisco}
\BIBentryALTinterwordspacing
Cisco, ``Cisco annual internet report (2018--2023) white paper,'' 2020.
  [Online]. Available:
  \url{https://www.cisco.com/c/en/us/solutions/collateral/executive-perspectives/annual-internet-report/white-paper-c11-741490.html.}
\BIBentrySTDinterwordspacing

\bibitem{wang2009mean}
Z.~Wang and A.~C. Bovik, ``Mean squared error: Love it or leave it? a new look
  at signal fidelity measures,'' \emph{IEEE signal processing magazine},
  vol.~26, no.~1, pp. 98--117, 2009.

\bibitem{toderici2015variable}
G.~Toderici, S.~M. O'Malley, S.~J. Hwang, D.~Vincent, D.~Minnen, S.~Baluja,
  M.~Covell, and R.~Sukthankar, ``Variable rate image compression with
  recurrent neural networks,'' \emph{arXiv preprint arXiv:1511.06085}, 2015.

\bibitem{toderici2017full}
G.~Toderici, D.~Vincent, N.~Johnston, S.~Jin~Hwang, D.~Minnen, J.~Shor, and
  M.~Covell, ``Full resolution image compression with recurrent neural
  networks,'' in \emph{IEEE Conference on Computer Vision and Pattern
  Recognition}, 2017, pp. 5306--5314.

\bibitem{balle2018variational}
J.~Ball{\'e}, D.~Minnen, S.~Singh, S.~J. Hwang, and N.~Johnston, ``Variational
  image compression with a scale hyperprior,'' in \emph{International
  Conference on Learning Representations}, 2018.

\bibitem{agustsson2017soft}
E.~Agustsson, F.~Mentzer, M.~Tschannen, L.~Cavigelli, R.~Timofte, L.~Benini,
  and L.~V. Gool, ``Soft-to-hard vector quantization for end-to-end learning
  compressible representations,'' in \emph{Advances in Neural Information
  Processing Systems}, 2017, pp. 1141--1151.

\bibitem{balle2019end}
J.~Ball{\'e}, V.~Laparra, and E.~Simoncelli, ``End-to-end optimized image
  compression,'' 2019.

\bibitem{johnston2018improved}
N.~Johnston, D.~Vincent, D.~Minnen, M.~Covell, S.~Singh, T.~Chinen,
  S.~Jin~Hwang, J.~Shor, and G.~Toderici, ``Improved lossy image compression
  with priming and spatially adaptive bit rates for recurrent networks,'' in
  \emph{IEEE Conference on Computer Vision and Pattern Recognition}, 2018, pp.
  4385--4393.

\bibitem{theis2017lossy}
L.~Theis, W.~Shi, A.~Cunningham, and F.~Husz{\'a}r, ``Lossy image compression
  with compressive autoencoders,'' \emph{International Conference on Learning
  Representations}, 2017.

\bibitem{rippel2017real}
O.~Rippel and L.~Bourdev, ``Real-time adaptive image compression,'' in
  \emph{International Conference on Machine Learning}, 2017, pp. 2922--2930.

\bibitem{agustsson2019generative}
E.~Agustsson, M.~Tschannen, F.~Mentzer, R.~Timofte, and L.~Van~Gool,
  ``Generative adversarial networks for extreme learned image compression,''
  pp. 221--231.

\bibitem{mentzer2018conditional}
F.~Mentzer, E.~Agustsson, M.~Tschannen, R.~Timofte, and L.~Van~Gool,
  ``Conditional probability models for deep image compression,'' pp.
  4394--4402, 2018.

\bibitem{minnen2018joint}
D.~Minnen, J.~Ball{\'e}, and G.~D. Toderici, ``Joint autoregressive and
  hierarchical priors for learned image compression,'' in \emph{Advances in
  Neural Information Processing Systems}, 2018, pp. 10\,771--10\,780.

\bibitem{patel2019deep}
Y.~Patel, S.~Appalaraju, and R.~Manmatha, ``Deep perceptual compression,''
  \emph{arXiv preprint arXiv:1907.08310}, 2019.

\bibitem{lee2018context}
J.~Lee, S.~Cho, and S.-K. Beack, ``Context-adaptive entropy model for
  end-to-end optimized image compression,'' \emph{arXiv preprint
  arXiv:1809.10452}, 2018.

\bibitem{blau2019rethinking}
Y.~Blau and T.~Michaeli, ``Rethinking lossy compression: The
  rate-distortion-perception tradeoff,'' in \emph{International Conference on
  Machine Learning}, 2019, pp. 675--685.

\bibitem{mentzer2019practical}
F.~Mentzer, E.~Agustsson, M.~Tschannen, R.~Timofte, and L.~Van~Gool,
  ``Practical full resolution learned lossless image compression,'' in
  \emph{2019 IEEE/CVF Conference on Computer Vision and Pattern Recognition
  (CVPR)}, 2019, pp. 10\,621--10\,630.

\bibitem{chen2017deepcoder}
T.~Chen, H.~Liu, Q.~Shen, T.~Yue, X.~Cao, and Z.~Ma, ``Deepcoder: A deep neural
  network based video compression,'' in \emph{2017 IEEE Visual Communications
  and Image Processing (VCIP)}.\hskip 1em plus 0.5em minus 0.4em\relax IEEE,
  2017, pp. 1--4.

\bibitem{wu2018video}
C.-Y. Wu, N.~Singhal, and P.~Krahenbuhl, ``Video compression through image
  interpolation,'' in \emph{Proceedings of the European Conference on Computer
  Vision (ECCV)}, 2018, pp. 416--431.

\bibitem{lu2019dvc}
G.~Lu, W.~Ouyang, D.~Xu, X.~Zhang, C.~Cai, and Z.~Gao, ``Dvc: An end-to-end
  deep video compression framework,'' in \emph{Proceedings of the IEEE
  Conference on Computer Vision and Pattern Recognition}, 2019, pp.
  11\,006--11\,015.

\bibitem{yang2020learning}
R.~Yang, F.~Mentzer, L.~Van~Gool, and R.~Timofte, ``Learning for video
  compression with hierarchical quality and recurrent enhancement,''
  \emph{arXiv preprint arXiv:2003.01966}, 2020.

\bibitem{soundararajan2012video}
R.~Soundararajan and A.~C. Bovik, ``Video quality assessment by reduced
  reference spatio-temporal entropic differencing,'' \emph{IEEE Transactions on
  Circuits and Systems for Video Technology}, vol.~23, no.~4, pp. 684--694,
  2012.

\bibitem{lee2021space}
D.~Y. Lee, H.~Ko, J.~Kim, and A.~C. Bovik, ``On the space-time statistics of
  motion pictures,'' \emph{JOSA A}, vol.~38, no.~7, pp. 908--923, 2021.

\bibitem{atick1990towards}
J.~J. Atick and A.~N. Redlich, ``Towards a theory of early visual processing,''
  \emph{Neural Computation}, vol.~2, no.~3, pp. 308--320, 1990.

\bibitem{attneave1954some}
F.~Attneave, ``Some informational aspects of visual perception.''
  \emph{Psychological review}, vol.~61, no.~3, p. 183, 1954.

\bibitem{dong1995temporal}
D.~W. Dong and J.~J. Atick, ``Temporal decorrelation: a theory of lagged and
  nonlagged responses in the lateral geniculate nucleus,'' \emph{Network:
  Computation in Neural Systems}, vol.~6, no.~2, pp. 159--178, 1995.

\bibitem{rucci2015unsteady}
M.~Rucci and J.~D. Victor, ``The unsteady eye: an information-processing stage,
  not a bug,'' \emph{Trends in Neurosciences}, vol.~38, no.~4, pp. 195--206,
  2015.

\bibitem{chichilnisky2002functional}
E.~Chichilnisky and R.~S. Kalmar, ``Functional asymmetries in on and off
  ganglion cells of primate retina,'' \emph{Journal of Neuroscience}, vol.~22,
  no.~7, pp. 2737--2747, 2002.

\bibitem{engbert2006microsaccades}
R.~Engbert, ``Microsaccades: A microcosm for research on oculomotor control,
  attention, and visual perception,'' \emph{Progress in Brain Research}, vol.
  154, pp. 177--192, 2006.

\bibitem{poletti2016compact}
M.~Poletti and M.~Rucci, ``A compact field guide to the study of microsaccades:
  Challenges and functions,'' \emph{Vision research}, vol. 118, pp. 83--97,
  2016.

\bibitem{olshausen1996emergence}
B.~A. Olshausen and D.~J. Field, ``Emergence of simple-cell receptive field
  properties by learning a sparse code for natural images,'' \emph{Nature},
  vol. 381, no. 6583, pp. 607--609, 1996.

\bibitem{wallace1992jpeg}
G.~K. Wallace, ``The {JPEG} still picture compression standard,'' \emph{IEEE
  Transactions on Consumer Electronics}, vol.~38, no.~1, pp. xviii--xxxiv,
  1992.

\bibitem{skodras2001jpeg}
A.~Skodras, C.~Christopoulos, and T.~Ebrahimi, ``The {JPEG} 2000 still image
  compression standard,'' \emph{IEEE Signal processing magazine}, vol.~18,
  no.~5, pp. 36--58, 2001.

\bibitem{bellard2015bpg}
\BIBentryALTinterwordspacing
F.~Bellard, ``{BPG} image format,'' vol.~1, 2015. [Online]. Available:
  \url{https://bellard.org/bpg}
\BIBentrySTDinterwordspacing

\bibitem{mukherjee2013latest}
D.~Mukherjee, J.~Bankoski, A.~Grange, J.~Han, J.~Koleszar, P.~Wilkins, Y.~Xu,
  and R.~Bultje, ``The latest open-source video codec vp9-an overview and
  preliminary results,'' in \emph{Picture Coding Symposium (PCS)}, 2013, pp.
  390--393.

\bibitem{lecun1995convolutional}
Y.~LeCun, Y.~Bengio \emph{et~al.}, ``Convolutional networks for images, speech,
  and time series,'' \emph{The Handbook of Brain Theory and Neural Networks},
  vol. 3361, no.~10, p. 1995, 1995.

\bibitem{hochreiter1997long}
S.~Hochreiter and J.~Schmidhuber, ``Long short-term memory,'' \emph{Neural
  Computation}, vol.~9, no.~8, pp. 1735--1780, 1997.

\bibitem{cho2014properties}
K.~Cho, B.~van Merri{\"e}nboer, D.~Bahdanau, and Y.~Bengio, ``On the properties
  of neural machine translation: Encoder--decoder approaches,'' \emph{Syntax,
  Semantics and Structure in Statistical Translation}, p. 103, 2014.

\bibitem{habibian2019video}
A.~Habibian, T.~v. Rozendaal, J.~M. Tomczak, and T.~S. Cohen, ``Video
  compression with rate-distortion autoencoders,'' \emph{IEEE International
  Conference on Computer Vision}, pp. 7033--7042, 2019.

\bibitem{rippel2019learned}
O.~Rippel, S.~Nair, C.~Lew, S.~Branson, A.~G. Anderson, and L.~Bourdev,
  ``Learned video compression,'' \emph{IEEE International Conference on
  Computer Vision}, pp. 3454--3463, 2019.

\bibitem{chen2019learning}
Z.~Chen, T.~He, X.~Jin, and F.~Wu, ``Learning for video compression,''
  \emph{IEEE Transactions on Circuits and Systems for Video Technology},
  vol.~30, no.~2, pp. 566--576, 2019.

\bibitem{cheng2019learning}
Z.~Cheng, H.~Sun, M.~Takeuchi, and J.~Katto, ``Learning image and video
  compression through spatial-temporal energy compaction,'' pp.
  10\,071--10\,080, 2019.

\bibitem{choi2017new}
G.~Choi, P.~Heo, S.~R. Oh, and H.~Park, ``A new motion estimation method for
  motion-compensated frame interpolation using a convolutional neural
  network,'' pp. 800--804, 2017.

\bibitem{choi2019deep}
H.~Choi and I.~V. Baji{\'c}, ``Deep frame prediction for video coding,''
  \emph{IEEE Transactions on Circuits and Systems for Video Technology}, 2019.

\bibitem{ranjan2017optical}
A.~Ranjan and M.~J. Black, ``Optical flow estimation using a spatial pyramid
  network,'' in \emph{Proceedings of the IEEE Conference on Computer Vision and
  Pattern Recognition}, 2017, pp. 4161--4170.

\bibitem{dosovitskiy2015flownet}
A.~Dosovitskiy, P.~Fischer, E.~Ilg, P.~Hausser, C.~Hazirbas, V.~Golkov, P.~Van
  Der~Smagt, D.~Cremers, and T.~Brox, ``Flownet: Learning optical flow with
  convolutional networks,'' pp. 2758--2766, 2015.

\bibitem{ilg2017flownet}
E.~Ilg, N.~Mayer, T.~Saikia, M.~Keuper, A.~Dosovitskiy, and T.~Brox, ``Flownet
  2.0: Evolution of optical flow estimation with deep networks,'' pp.
  2462--2470, 2017.

\bibitem{hui2018liteflownet}
T.-W. Hui, X.~Tang, and C.~Change~Loy, ``Liteflownet: A lightweight
  convolutional neural network for optical flow estimation,'' pp. 8981--8989,
  2018.

\bibitem{xingjian2015convolutional}
S.~Xingjian, Z.~Chen, H.~Wang, D.-Y. Yeung, W.-K. Wong, and W.-c. Woo,
  ``Convolutional lstm network: A machine learning approach for precipitation
  nowcasting,'' in \emph{Advances in neural information processing systems},
  2015, pp. 802--810.

\bibitem{kay2017kinetics}
W.~Kay, J.~Carreira, K.~Simonyan, B.~Zhang, C.~Hillier, S.~Vijayanarasimhan,
  F.~Viola, T.~Green, T.~Back, P.~Natsev \emph{et~al.}, ``The kinetics human
  action video dataset,'' \emph{arXiv preprint arXiv:1705.06950}, 2017.

\bibitem{carreira2018short}
J.~Carreira, E.~Noland, A.~Banki-Horvath, C.~Hillier, and A.~Zisserman, ``A
  short note about kinetics-600,'' \emph{arXiv preprint arXiv:1808.01340},
  2018.

\bibitem{xue2019video}
T.~Xue, B.~Chen, J.~Wu, D.~Wei, and W.~T. Freeman, ``Video enhancement with
  task-oriented flow,'' \emph{International Journal of Computer Vision}, vol.
  127, no.~8, pp. 1106--1125, 2019.

\bibitem{vtl}
\BIBentryALTinterwordspacing
V.~T. Library, ``{VTL} test sequences,'' 2020. [Online]. Available:
  \url{http://trace.eas.asu.edu/index.html.}
\BIBentrySTDinterwordspacing

\bibitem{bossen2013common}
F.~Bossen \emph{et~al.}, ``Common test conditions and software reference
  configurations,'' \emph{JCTVC-L1100}, vol.~12, p.~7, 2013.

\bibitem{uvg}
\BIBentryALTinterwordspacing
{Ultra Video Group}, ``{UVG} test sequences,'' 2020. [Online]. Available:
  \url{http://ultravideo.cs.tut.fi/#testsequences.}
\BIBentrySTDinterwordspacing

\bibitem{lu2020end}
G.~Lu, X.~Zhang, W.~Ouyang, L.~Chen, Z.~Gao, and D.~Xu, ``An end-to-end
  learning framework for video compression,'' \emph{IEEE Transactions on
  Pattern Analysis and Machine Intelligence}, 2020.

\bibitem{wang2003multiscale}
Z.~Wang, E.~P. Simoncelli, and A.~C. Bovik, ``Multiscale structural similarity
  for image quality assessment,'' vol.~2, pp. 1398--1402, Nov. 2003.

\bibitem{wiegand2003overview}
T.~Wiegand, G.~J. Sullivan, G.~Bjontegaard, and A.~Luthra, ``Overview of the h.
  264/avc video coding standard,'' \emph{IEEE Transactions on Circuits and
  Systems for Video Technology}, vol.~13, no.~7, pp. 560--576, 2003.

\bibitem{sullivan2012overview}
G.~J. Sullivan, J.-R. Ohm, W.-J. Han, and T.~Wiegand, ``Overview of the high
  efficiency video coding {(HEVC)} standard,'' \emph{IEEE Transactions on
  Circuits and Systems for Video Technology}, vol.~22, no.~12, pp. 1649--1668,
  2012.

\bibitem{bross2021overview}
B.~Bross, Y.-K. Wang, Y.~Ye, S.~Liu, J.~Chen, G.~J. Sullivan, and J.-R. Ohm,
  ``Overview of the versatile video coding (vvc) standard and its
  applications,'' \emph{IEEE Transactions on Circuits and Systems for Video
  Technology}, vol.~31, no.~10, pp. 3736--3764, 2021.

\bibitem{h266}
\BIBentryALTinterwordspacing
{Fraunhofer Heinrich Hertz Institute}, ``{Fraunhofer Versatile Video Encoder
  (VVenC)},'' 2021. [Online]. Available:
  \url{https://www.hhi.fraunhofer.de/en/departments/vca/technologies-and-solutions/h266-vvc.html.}
\BIBentrySTDinterwordspacing

\bibitem{brites2015distributed}
C.~Brites and F.~Pereira, ``Distributed video coding: Assessing the hevc
  upgrade,'' \emph{Signal Processing: Image Communication}, vol.~32, pp.
  81--105, 2015.

\end{thebibliography}

% \end{IEEEbiography}



\end{document}